\documentclass[a4paper,fleqn,usenatbib]{mnras}

\usepackage{lineno}
\usepackage[T1]{fontenc}
\usepackage{ae,aecompl}
\usepackage{textcomp}
\usepackage{graphicx}	
\usepackage{amsmath}	
\usepackage{amssymb}	
\usepackage{dirtytalk}
\usepackage{siunitx}
\usepackage{tablefootnote}

\usepackage{xcolor}

\title[SN2016iyc ]{SN 2016iyc: A Type IIb supernova arising from a low-mass progenitor}

\author[Aryan et al.]
{Amar Aryan\thanks{e-mail :amar@aries.res.in, amararyan941@gmail.com}$^{1,2}$,
S. B. Pandey$^{1}$, WeiKang Zheng$^{3}$, Alexei V. Filippenko$^{3,4}$, Jozsef Vinko$^{5,6,7,8}$,
\newauthor
 Ryoma Ouchi$^{9}$, Thomas G. Brink$^{3}$, Andrew Halle$^{3}$, Jeffrey Molloy$^{3}$, Sahana Kumar$^{3,10}$,
\newauthor
 Goni Halevi$^{3,11}$, Charles D. Kilpatrick$^{12}$, Amit Kumar$^{1,13}$, Rahul Gupta$^{1,2}$, and Amit Kumar Ror$^{1}$
\\\\
$^{1}$Aryabhatta Research Institute of Observational Sciences, Manora Peak, Nainital 263002, India\\
$^{2}$Department of Physics, Deen Dayal Upadhyay Gorakhpur University, Gorakhpur (U.P.) 273009, India\\
$^{3}$Department of Astronomy, University of California, Berkeley, CA 94720-3411, USA\\
$^{4}$Miller Institute for Basic Research in Science, University of California,
Berkeley, CA 94720, USA\\
$^{5}$Department of Astronomy, University of Texas at Austin, Austin, TX 78712, USA \\
$^{6}$CSFK Konkoly Observatory, Konkoly Thege M. ut 15-17, Budapest, 1121, Hungary\\
$^{7}$ELTE E\"otv\"os Lor\'and University, Institute of Physics, P\'azm\'any P\'eter s\'et\'any 1/A, Budapest, 1117 Hungary\\
$^{8}$Department of Optics and Quantum Electronics, University of Szeged, D\'om t\'er 9, Szeged, 6720 Hungary\\
$^{9}$Visual Intelligence Laboratories, NEC Corporation, Kanagawa 211-8666, Japan\\
$^{10}$Department of Physics, Florida State University, Tallahassee, FL 32306,
USA\\
$^{11}$Department of Astrophysical Sciences, Princeton University, 4 Ivy Lane,
Princeton, NJ 08540, USA\\
$^{12}$Center for Interdisciplinary Exploration and Research in Astrophysics and Department of Physics and Astronomy,\\ $    $ Northwestern University, 1800 Sherman Avenue, 8th Floor, Evanston, IL 60201, USA\\
$^{13}$School of Studies in Physics and Astrophysics, Pt. Ravishankar Shukla University, Chattisgarh 492010, India\\
}

\date{Accepted XXX. Received YYY; in original form ZZZ}

\pubyear{2022}

\begin{document}
\label{firstpage}
\pagerange{\pageref{firstpage}--\pageref{lastpage}}
\maketitle

\begin{abstract}
In this work, photometric and spectroscopic analyses of a very low-luminosity Type IIb supernova (SN) 2016iyc have been performed. SN~2016iyc lies near the faint end among the distribution of similar supernovae (SNe). 
Given lower ejecta mass ($M_{\rm ej}$) and low nickel mass ($M_{\rm Ni}$) from the literature, combined with SN~2016iyc lying near the faint end, one-dimensional stellar evolution models of 9--14\,M$_{\odot}$ zero-age main-sequence (ZAMS) stars as the possible progenitors of SN~2016iyc have been performed using the publicly available code {\tt MESA}. Moreover, synthetic explosions of the progenitor models have been simulated using the hydrodynamic evolution codes {\tt STELLA} and {\tt SNEC}. The bolometric luminosity light curve and photospheric velocities produced through synthetic explosions of ZAMS stars of mass in the range 12--13\,M$_{\odot}$ having a pre-supernova radius $R_{\mathrm{0}} =$ (240--300)\,R$_{\odot}$, with $M_{\rm ej} =$ (1.89--1.93)\,M$_{\odot}$, explosion energy $E_{\rm exp} = $ (0.28--0.35) $\times 10^{51}$\,erg, and $M_{\rm Ni} < 0.09$\,M$_{\odot}$, are in good agreement with observations; thus, SN~2016iyc probably exploded from a progenitor near the lower mass limits for SNe~IIb. Finally, hydrodynamic simulations of the explosions of SN~2016gkg and SN~2011fu have also been performed to compare intermediate- and high-luminosity examples among well-studied SNe~IIb. The results of progenitor modelling and synthetic explosions for SN~2016iyc, SN~2016gkg, and SN~2011fu exhibit a diverse range of mass for the possible progenitors of SNe~IIb.

\end{abstract}


\section{Introduction}
\label{sec:Introduction}

Type IIb supernovae (SNe) are a subclass of catastrophic core-collapse SNe (CCSNe). These SNe form a transition class of objects that link hydrogen (H)-rich Type II and H-deficient Type Ib SNe \citep[][]{Filippenko1988, Filippenko1993, Smartt2009}; see \citet{Filippenko1997} for a review. Their early-phase spectra show strong H features, and distinct helium (He) lines start to appear a few weeks later; thus, these SNe are thought to be partially stripped by retaining a significant H envelope, and the He core is exposed once the envelope becomes optically thin.

The predominant powering mechanisms in SNe~IIb are the radioactive decay of $^{56}$Ni and the deposition of internal energy by the shock in the ejecta \citep[e.g.,][]{Arnett1980, Arnett1982, Arnett1996, Nadyozhin1994, Chatzopoulous2013, Nicholl2017}. In a few cases, the SN progenitors are also surrounded by dense circumstellar material (CSM) which may interact violently with the SN ejecta. The interaction of CSM with the SN ejecta results in the formation of a two-component shock structure: a forward shock moving into the CSM and a reverse shock moving back into the SN ejecta. Both of these shocks deposit their kinetic energies into the material that is radiatively released, powering the light curves of the SNe \citep[e.g.,][]{ Chevalier1982, Chevalier1994, Moriya2011, Ginzberg2012, Chatzopoulous2013, Nicholl2017}.

Understanding the possible progenitors of stripped or partially-stripped CCSNe is a challenging task. Methods to investigate the SN progenitors and their properties include (a) direct detections of objects in pre-explosion images, and (b) modelling of certain mass zero-age main-sequence (ZAMS) stars as the possible progenitors, based on the observed photometric and spectroscopic properties of the SNe. Direct detections of progenitors are rare owing to the uncertainty associated with the spatial positions and the infrequent occurrence of these transient phenomena. One has to be very lucky to get such pre-explosion images. However, for SNe~IIb, four cases of the direct detection of objects in pre-explosion images have been claimed. These include SN~1993J \citep[][]{Filippenko1993-IAUC,Aldering1994}, SN~2008ax \citep[][]{Crockett2008}, SN~2011dh \citep[][]{Maund2011,Van2011}, and SN~2013df \citep[][]{Van2014}, indicating either massive Wolf-Rayet (WR) stars ($M_{\rm ZAMS} \approx 10$--28\,M$_\odot$; \citealt{Crockett2008}) or more extended yellow supergiants (YSGs) with $M_{\rm ZAMS} = 12$--17\,M$_\odot$ \citep[][]{Van2013, Folatelli2014, Smartt2015} as SN~IIb progenitors. Following \citet[][]{Smartt2009b} and \citet[][]{Van2017}, there have only been $\sim 34$ cases of direct CCSNe progenitor detections. With these direct detections, the progenitors of SNe~IIP are red supergiants (RSGs); SN~IIn progenitors are luminous blue variables; the progenitors of SNe~IIL are still debated, with only the case of SN~2009kr suggesting RSG or yellow supergiant progenitors; and the progenitors of SNe~Ib/c are either low-mass stars in a binary system \citep[][]{Podsiadlowski1992, Nomoto1995, Smartt2009} or a single massive WR star \citep[e.g.,][]{Gaskell1986, Eldridge2011, Groh2013}.  

The second method, progenitor modelling using stellar evolution codes to constrain the nature of the possible progenitors of stripped or partially-stripped CCSNe, identified either via direct imaging as in the case of iPTF13bvn \citep[][]{Cao2013} or indirect methods including nebular-phase spectral modelling \citep[][]{Jerkstrand2015, Uomoto1986}, and simulating the synthetic explosions of their pre-SN models, is also vital to understand their nature, physical conditions, circumstellar environment, and chemical compositions. But, the progenitor modelling of such objects using various stellar evolution codes is difficult owing to the complicated stages of shell burning. Another problem associated with such modelling is the obscure nature of the mixing-length-theory parameter ($\alpha_{\rm MLT}$). The basis of $\alpha_{\rm MLT}$ has no physical origin \citep[][]{Joyce2018,Viani2018}. Furthermore, \citet[][]{Joyce2018} mentions that $\alpha_{\rm MLT}$ is neither a physical constant nor a computational one; it is rather a free parameter, so the value of $\alpha_{\rm MLT}$ must be determined individually in each stellar evolution code.

Owing to the above-mentioned difficulties, only a handful of such studies including progenitor modelling followed by their synthetic explosions have been performed in the case of stripped or partially-stripped CCSNe, including the Type Ib SN~iPTF13bvn \citep[][]{Cao2013, Bersten2014, Paxton2018}, the famous Type IIb SN~2016gkg \citep[][]{Bersten2018}, a few other Type IIb SNe including SN~2011dh \citep[][]{Bersten2012}, SN~2011fu \citep[][]{Morales2015}, two Type Ib SNe~2015ap and 2016bau \citep[][]{Aryan2021}, and another Type Ib SN~2012au \citep[][]{Pandey2021}.

Considering these limited studies, our work goes one step further, as we perform the one-dimensional stellar evolution of the possible progenitors of the low-luminosity Type IIb SN~2016iyc and also simulate synthetic explosions. Our studies in this work point toward SN~2016iyc originating from the lower-mass end of the ZAMS progenitor systems observed for Type IIb CCSNe.

This paper is divided into eight sections, including an introduction in Sec.~\ref{sec:Introduction}. Sec.~\ref{sec:Data_red} provides details about various telescopes and reduction procedures, including the discovery of SN~2016iyc using the Katzman Automatic Imaging Telescope (KAIT) at Lick Observatory as well as recalibrated photometry of SN~2016gkg. In Sec.~\ref{sec:Photometric}, methods to correct for the extinction, photometric properties including the bolometric light curve, black-body temperature, and radius evolutions are discussed. We present the analyses describing the spectral properties and comparisons with other similar and well-studied SNe in Sec.~\ref{sec:Spectral}; we also model the spectra of these SNe using {\tt SYN++}. The assumptions and methods for modelling the possible progenitor of SN~2016iyc and the evolution of the models until the onset of core collapse using {\tt MESA} are presented in  Sec.~\ref{sec:mesa_snec}. Further, in this section, we discuss the assumptions and methods for simulating the synthetic explosions using {\tt SNEC} and {\tt STELLA}. Here, comparisons between the parameters obtained through synthetic explosions and observed ones are presented. We also perform hydrodynamic modelling of the synthetic explosions of SN~2016gkg and SN~2011fu in Sec.~\ref{sec:SN2016gkg_model}. In Sec.~\ref{sec:Discussions}, we discuss our major results and findings. We summarise our work and provide concluding remarks in Sec.~\ref{sec:Conclusions}. 

\section{Data acquisition and reduction }
\label{sec:Data_red}
SN~2016iyc was discovered \citep[][]{de2016} in an 18\,s unfiltered image taken at 03:28:00 on 2016~Dec.~18 (UT dates are used throughout this paper) by the 0.76\,m KAIT as part of the Lick Observatory Supernova Search \citep[LOSS;][]{Filippenko2001}. Its brightness was $17.81\pm0.11$\,mag, and the object was not detected earlier on Dec. 04.14 with an upper limit of 19.0\,mag.
We measure its J2000.0 coordinates to be $\alpha=22^{\mathrm{h}}09^{\mathrm{m}}14\farcs29$,
$\delta=+21^{\circ}31\arcmin17\farcs3$, with an uncertainty of $0\farcs5$ in each coordinate.
SN~2016iyc is $14\farcs0$ west and $10\farcs4$ north of the nucleus of the host galaxy UGC~11924, which has
redshift $z=0.012685 \pm 0.000017$ \citep[][]{Giovanelli1993}, with a spiral morphology (Sd).
$B$, $V$, $R$, and $I$ multiband follow-up images of SN~2016iyc were obtained with both KAIT and the 1\,m Nickel telescope at Lick Observatory; KAIT also obtained additional unfiltered ({\it Clear (C)}-band) images. Although unfiltered and thus nonstandard, $C$ is most similar to the $R$ band \citep[][]{Li2003}, and has been widely used for SN observations by KAIT \citep[e.g.,][]{Stahl2019,dejager2019,Zheng2022}.

All images were reduced using a custom pipeline\footnote{https://github.com/benstahl92/LOSSPhotPypeline}
detailed by \citet[][]{Stahl2019}. Here we briefly summarise the photometric procedure. Image subtraction was conducted in order to remove the host-galaxy contribution, using additional images obtained after the SN had faded below the detection limit. Point-spread-function (PSF) photometry was obtained using DAOPHOT \citep[][]{Stetson1987} from the IDL Astronomy User’s Library\footnote{http://idlastro.gsfc.nasa.gov/}.
Several nearby stars were chosen from the Pan-STARRS1\footnote{http://archive.stsci.edu/panstarrs/search.php} catalogue for calibration purpose;
their magnitudes were first transformed into the \citet{Landolt1992} system using the empirical prescription presented by \citet[][Eq.~6]{Torny2012}, and then into the KAIT/Nickel natural system.
All apparent magnitudes were measured in the KAIT4/Nickel2 natural system. The final results were transformed to the standard system using local calibrators and colour terms for KAIT4 and Nickel2 \citep[][]{Stahl2019}.

\begin{figure*}
\centering
    \includegraphics[height=8.0cm,width=8.5cm,angle=0]{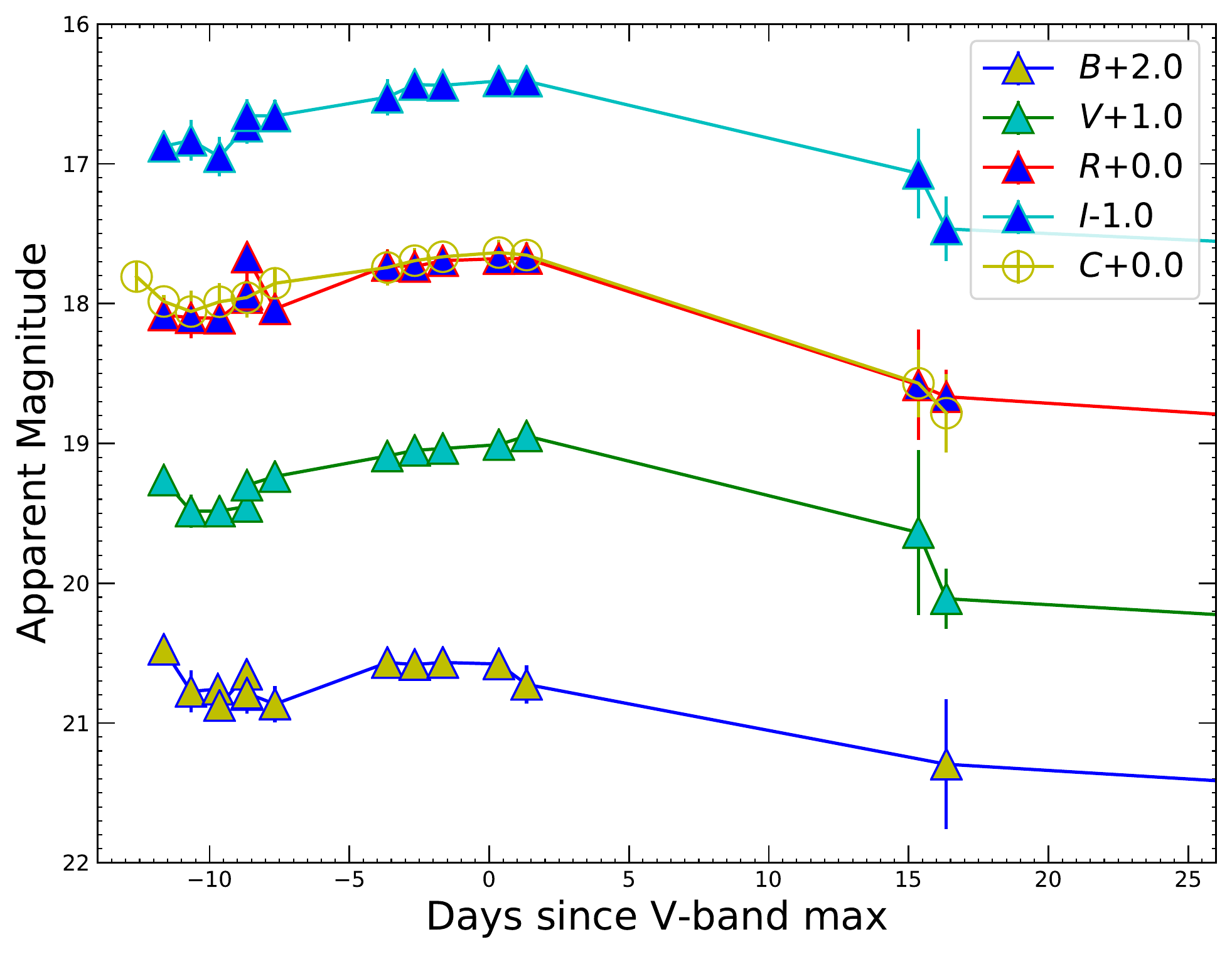}
    \includegraphics[height=8.0cm,width=8.5cm,angle=0]{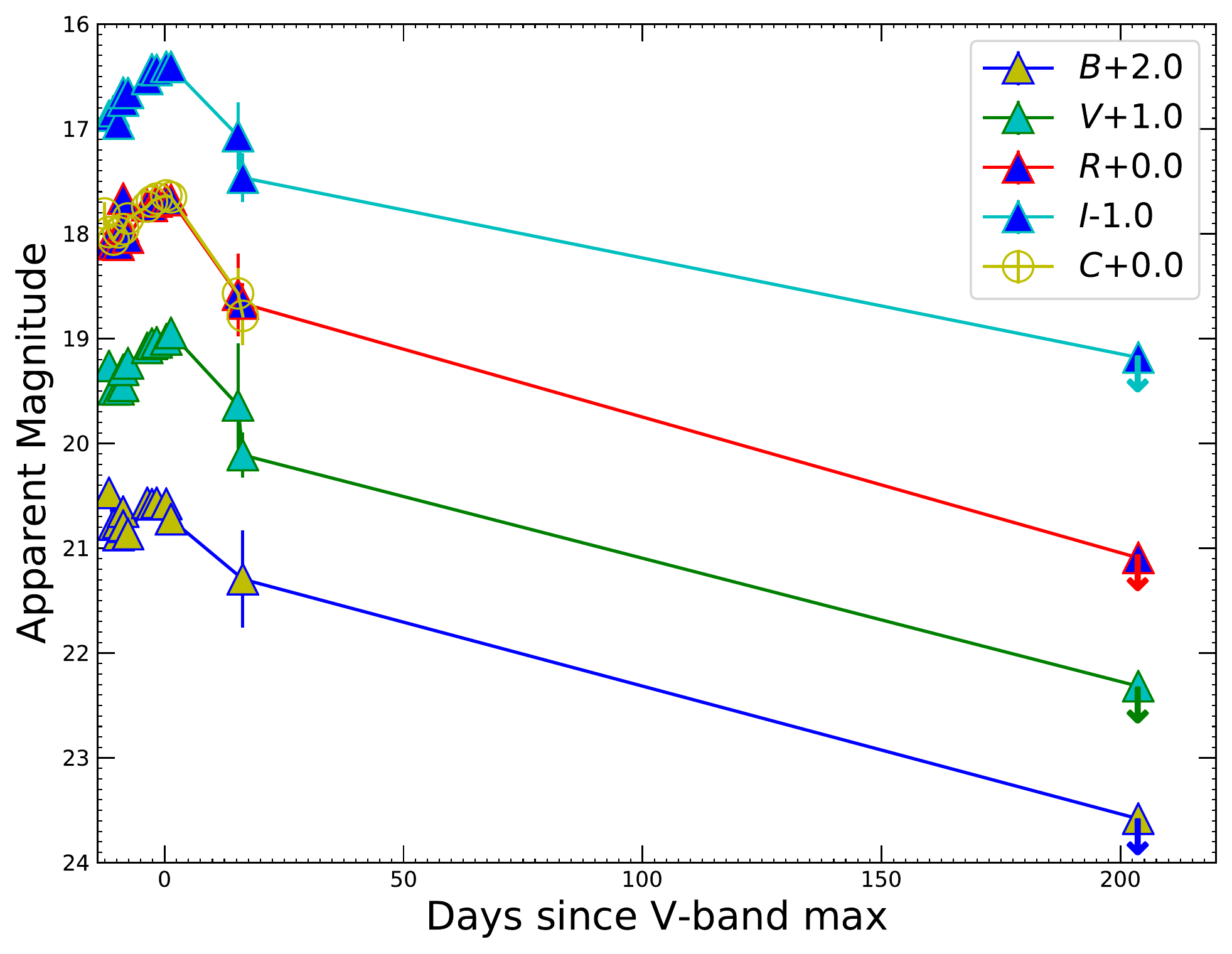}
   \caption{ {\em Left:}  The $BVRI$- and $C$-band light curves of SN~2016iyc, obtained with KAIT. The generic extended-SBO feature of SNe~IIb is visible in each band. {\em Right:} The $BVRI$ and $C$ light curves along with upper limits in each band using the Las Cumbres Observatory global telescope network. The upper limits in the last epoch are extremely useful for setting an upper limit on  $M_{\rm Ni}$.}
    \label{fig:BVRIC_LC}
\end{figure*}

The same method was adopted to reprocess the LOSS data of SN~2016gkg (originally published by \citealt{Bersten2018}), except that no subtraction procedure was applied to SN~2016gkg; the calibration source was also chosen from the Pan-STARRS1 catalog. Photometry of SN~2016gkg at two epochs was also obtained with the 3.6\,m Devasthal optical  telescope (DOT) using the 4K$\times$4K CCD Imager \citep{Pandey2018, Kumar2021a}. SN~2016gkg was the first SN detected by the 3.6\,m DOT during its initial commissioning phases. For the data obtained from the 3.6\,m DOT, the \citet{Landolt1992} photometric standard fields PG 0918, PG 1633, and PG 1657 were observed on 2021 Feb. 07 along with the SN field in the $UBVRI$ bands under good photometric conditions. These three Landolt fields have standard stars with a $V$-band magnitude range of 12.27--15.26\,mag and a $B-V$ colour range of $-$0.27 to +1.13\,mag. The SN fields observed in 2021 were used for template subtraction to remove the host-galaxy contributions from the source images. Template subtraction was performed with standard procedures by matching the full width at half-maximum intensity (FWHM) and flux values of respective images. The optical photometric data reduction and calibration were made with a standard process discussed by \citet[][]{Kumar2021} and Python scripts hosted on \textsc{RedPipe} \citep[][]{Singh2021}. The average atmospheric extinction values in the $U$, $B$, $V$, $R$, and $I$ bands for the Devasthal site were adopted from \citet{Kumar2021a}. The recalibrated KAIT data of SN~2016gkg along with those observed at later epochs using the 4K$\times$4K CCD Imager mounted at the axial port of the 3.6\,m DOT were used for the construction of bolometric light curves as described in the following sections.

A single optical spectrum of SN~2016iyc was obtained on 2016 Dec. 23 with the Kast double spectrograph \citep[][]{MillerStone1993} mounted on the 3\,m Shane telescope at Lick Observatory.  The 2700\,s exposure was taken at or near the parallactic angle to minimise slit losses caused by atmospheric dispersion \citep[][]{Filippenko1982}. The observations were conducted with a $2''$-wide slit, 600/4310 grism on the blue side, and 300/7500 grating on the red side.  This instrument configuration has a combined wavelength range of $\sim 3500$--10,400\,\si{\angstrom} and spectral resolving power of $R \approx 800$.  Data reduction followed standard techniques for CCD processing and spectrum extraction \citep[][]{Silverman2012} utilising IRAF\footnote{IRAF is distributed by the National Optical Astronomy Observatory, which is operated by AURA, Inc., under a cooperative agreement with the U.S. NSF.} routines and custom Python and IDL codes\footnote{https://github.com/ishivvers/TheKastShiv}.  Low-order polynomial fits to comparison-lamp spectra were used to calibrate the wavelength scale, and small adjustments derived from night-sky lines in the target frames were applied.  Observations of appropriate spectrophotometric standard stars were used to flux calibrate the spectrum.

\section{Photometric Properties}
\label{sec:Photometric}
In this section, we discuss the photometric properties of SN~2016iyc, including the colour evolution, extinction, bolometric light curves, and various black-body parameters. The $BVRI$- and $C$-band photometric data of SN~2016iyc are presented in Table~\ref{tab:optical_observations_2016iyc}.

\begin{figure}	
    \includegraphics[width=\columnwidth]{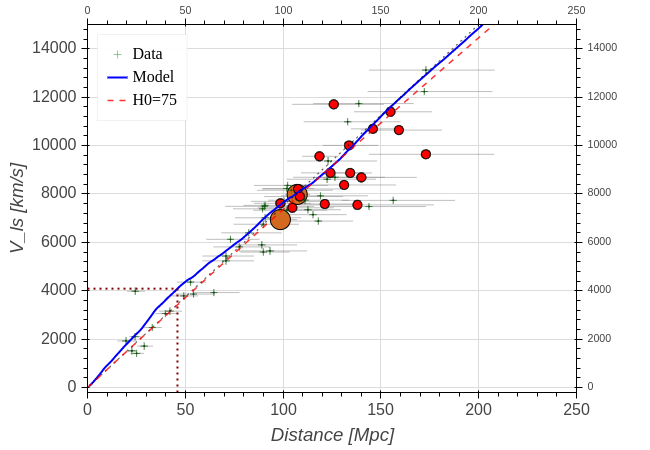}
    \caption{ Distance estimation for the nearby ($z=0.012685$) SN~2016iyc field ($\alpha=22^{\mathrm{h}}09^{\mathrm{m}}14\farcs29$,
$\delta=+21^{\circ}31\arcmin17\farcs3$) using the method described by \citet[][]{Kourkchi2020} with \citet[][]{Graziani2019} models. The distance value estimated with this method is $\sim 10$\,Mpc ($\sim 20$\%) nearer than that reported using the published redshift value \citep[][]{Planck2016} for SN~2016iyc.}
    \label{fig:distance}
\end{figure}

Most of the analyses in this paper have been performed with respect to the phase of $V$-band maximum brightness.  The photometric data of SN~2016iyc lack dense coverage near peak brightness; thus, to find the phase of $V$-band maximum, we used the $V$-band light curve of SN~2013df as a template having a rising timescale similar to that of SN~2016iyc (Figure~\ref{fig:V_max}). We fit a fourth-order polynomial to the template light curve and find the date of $V$ maximum to be MJD $57752.7 \pm 0.2$. The left panel of Figure~\ref{fig:BVRIC_LC} shows the $BVRI$- and $C$-band light curves of SN~2016iyc. The characteristic extended shock-breakout (hereafter, extended-SBO) feature typically observed in SNe~IIb is seen in all of the bands. Multiple mechanisms and/or ejecta/progenitor properties have been theorised to explain such enhancement in the luminosity before the primary peak, including an increase in the progenitor radius up to a few 100\,R$_{\odot}$ \citep[e.g.][]{Nomoto1993,Podsiadlowski1993,Woosley1994}; an interaction with CSM similar to the case of Type IIn SNe \citep[][]{Schlegel1990}; in a close-binary system, the interaction with the companion \citep[][]{Kasen2010,Moriya2015}; and sometimes enhanced $^{56}$Ni mixing into the outer ejecta \citep[e.g.,][]{Arnett1989}. The right-hand panel of Figure~\ref{fig:BVRIC_LC} shows the $BVRI$ and $C$ light curves along with the late-time upper limits in each band. These upper limits are very useful in constraining the upper limit on $M_{\rm Ni}$.

\subsection{Distance estimation of SN~2016iyc}

\begin{table*}
  \caption {The adopted total extinction values, distances, and corresponding distance moduli of a subset of SNe considered here.}
\label{tab:comparison_Sample}
\begin{center}
{\scriptsize
\begin{tabular}{ccccccccccccc}
\hline\hline
	SN name    &	$E(B-V)_{\rm tot}$	&	Adopted distance  & Distance modulus & $V_{\rm max}$ & log\,$(L_{BVRI})_p$ \\
               &	(mag)	&	(Mpc) & (mag) & (mag) & (erg\,s$^{-1}$) \\
\hline
SN~1987A  	 	&	0.16 \citep[][]{Bose2021}	&	0.05 \citep[][]{Bose2021} & 18.44 & $-15.52 \pm 0.02$ & 42.555 $\pm$ 0.004\\\\

SN~1993J  	 	&	0.18 \citep[][]{Richmond1996}	&	3.68 \citep[][]{Bose2021} & 27.82 & $-16.97 \pm 0.03$ & 42.01 $\pm$ 0.01\\\\

SN~2003bg  	 	&	0.02 \citep[][]{Mazzali2009}	&	24 \citep[][]{Mazzali2009}   & 31.90 & -17.8$\pm$0.2 & 42.31 $\pm$ 0.03\\\\

SN~2008ax  	    &	0.3 \citep[][]{Tsvetkov2009}	&	 9.6 \citep[][]{Pastorello2008}   & 29.92 & $-16.35 \pm 0.05$   & 42.07  $\pm$ 0.03\\\\

SN~2011dh  	 	&	0.035 \citep[][]{Sahu2013}	&	 8.4 \citep[][]{Sahu2013}   & 29.62 & $-17.06 \pm 0.02$ & 41.99 $\pm$ 0.03\\\\

SN~2011fu 	 	&	0.22 \citep[][]{Kumar2013}	&	 77.0 \citep[][]{Kumar2013}   & 34.46 & $-17.51 \pm 0.03$ & 42.426 $\pm$ 0.006\\\\

SN~2011hs  	 	&	0.17 \citep[][]{Bufano2014}	&	 23.44$^{*}$     & 31.85 & $-16.03 \pm 0.03$ & 41.74 $\pm$ 0.02\\\\

SN~2013df  	 	&	0.098 \citep[][]{Morales2015}	&	 16.6 \citep[][]{Van2014}   & 31.1 & $-16.47 \pm 0.05$ & 41.87 $\pm$ 0.01\\\\

SN~2016gkg 	&	0.017 \citep[][]{Bersten2018}	&	 26.4 \citep[][]{Kilpatrick2017}   & 32.11 & $-17.03 \pm 0.05$ & 41.98 $\pm$ 0.02\\\\

SN~2016iyc 	&	0.137 	&	 46.0   &  33.31 & $-15.32 \pm 0.05$ & 41.44 $\pm$ 0.01\\

\hline\hline
\end{tabular}}
\end{center}
{The sources for the $BVRI$ light curves for SNe in the comparison sample are as follows. SN~1987A, \citet[][]{Menzies1987} and \citet[][]{Makino1987}; SN~1993J, \citet[][]{Zhang2004}; SN~2003bg, \citet[][]{Hamuy2009}; SN~2008ax, \citet[][]{Pastorello2008}; SN~2011dh, \citet[][]{Sahu2013}; SN~2011fu, \citet[][]{Kumar2013}; SN~2011hs, \citet[][]{Bufano2014}; SN~2013df, \citet[][]{Morales2015}. Adopted distances have been used to calculate the distance moduli. The total extinction correction and distance moduli for all the SNe in the comparison sample have been taken into account while calculating the bolometric light curves. $^*$For SN~2011hs, the distance modulus is 31.85\,mag \citep[][]{Bufano2014}, which is used to back-calculate a distance of 23.44\,Mpc.}
\end{table*}

Distance determinations from redshifts ($z$) are severely biased for nearby SNe because of the peculiar motions of nearby galaxies that are comparable to the Hubble flow. So, the redshift-based distance estimates can be used only for SNe having $z > 0.1$. Hence, the redshift-based distance for SN~2016iyc, published by \citet[][]{Planck2016}, could be spurious. SN~2016iyc being nearby ($z = 0.012685$), we cross-verified the redshift-based distance estimate (56.6837\,Mpc as mentioned by \citealt[][]{Planck2016}) with an advanced tool (Figure~\ref{fig:distance}) recently featured by \citet[][]{Kourkchi2020}\footnote{http://edd.ifa.hawaii.edu/CF3calculator/}, known as the Distance-Velocity ($D$--$V$) Calculator. Corresponding to a heliocentric velocity $V_{\rm h} \approx 3804$\,km\,s$^{-1}$, the observed velocity ($V_{\rm ls}$) at the location of SN~2016iyc is found to be $\sim 4089$\,km\,s$^{-1}$ by utilising Eq.~5 of \citet[][]{Kourkchi2020}. Corresponding to $V_{\rm ls} = 4089$\,km\,s$^{-1}$, the $D$--$V$ Calculator gives a distance of $\sim 46$\,Mpc, which is $\sim 20$\% less than  \citet[][]{Planck2016}. The distance modulus for SN~2016iyc corresponding to this distance is 33.31\,mag and is adopted for all further analyses in this paper. The distances for the SNe used as a comparison sample are well established in the literature and are used as such for the estimation of their respective bolometric luminosities. The distance of each SN in the comparison sample along with the corresponding distance modulus is presented in Table~\ref{tab:comparison_Sample}.

\subsection{Colour evolution and extinction correction}
For SN~2016iyc, we corrected for the Milky Way (MW) extinction using NED, following \citet[][]{Schlafly2011}. In the direction of SN~2016iyc, $E(B-V)_{\rm MW} = 0.067$\,mag, so the MW extinction corrections for the $B$, $V$, $R$, and $I$ bands are 0.278, 0.207, 0.155, and 0.099\,mag, respectively.

\begin{figure}	
    \includegraphics[width=\columnwidth]{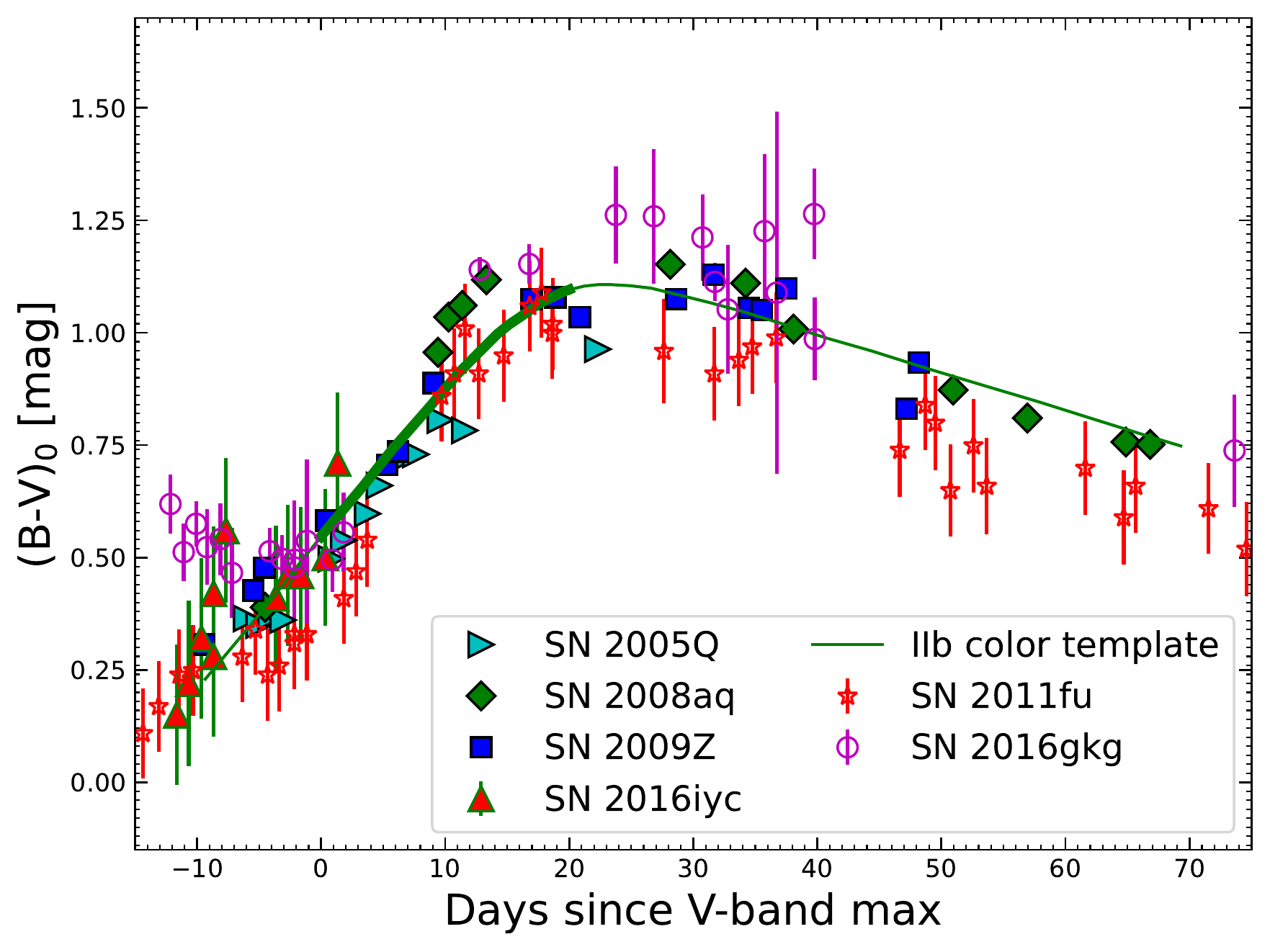}
    \caption{ The total-extinction-corrected $(B-V)_{0}$ colour curves of SN~2016iyc, SN~2011fu, and SN~2016gkg, plotted along with other SNe~IIb. The data for SN~2005Q, SN~2008aq, and SN~2009Z were taken from \citet[][]{Stritzinger2018}, with these three SNe analysed to have negligible host-galaxy extinction. The green curve shows the template $(B-V)_{0}$ curve for SNe~IIb having negligible host-galaxy extinction. The thick portion of the template curve shows the 0 to +20\,d period that should be considered when determining the colour excess for the reasons mentioned by \citet[][]{Stritzinger2018}.}
    \label{fig:color_curve}
\end{figure}

Only one spectrum of SN~2016iyc is available, and it does not exhibit a clear Na~I~D absorption line produced by gas in the host galaxy, suggesting that there is negligible host-galaxy extinction. However, neglecting host-galaxy extinction based on only the absence of obvious Na~I~D could be spurious. 
In a recently published Lick/KAIT data-release paper for various stripped-envelope SNe, \citet[][]{Zheng2022} performed comprehensive analysis to determine the host-galaxy contamination and found $E(B-V)_{\rm host} = 0.07$\,mag for SN~2016iyc. We also performed a simple analysis to put an upper limit on the host-galaxy extinction. Five early epochs were selected, and the spectral energy distribution (SED) was fitted with black-body curves by assuming different $E(B-V)_{\rm host}$ values (Figure~\ref{fig:extinction_trial}). We found that going beyond 0.07\,mag of host-galaxy extinction results in black-body temperature exceeding 11,200\,K. Such high temperatures are generally not seen in SNe~IIb. Following \citet[][]{Ben2015}, the early-time black-body temperatures associated with SN~1993J, SN~2011dh, and SN~2013df are 8200\,K, 8200\,K, and 7470\,K, respectively. There have been only a few cases where the early black-body temperature exceeds 11,000\,K; one such example is SN~2001ig \citet[][]{Ben2015}, but this SN may have come from a compact WR binary progenitor system\citep[][]{Ryder2004}.

Based on the above analyses and the results of \citet[][]{Zheng2022}, we adopt a host-galaxy extinction of 0.07\,mag throughout this paper. Thus, a total (Milky Way + host-galaxy) extinction of $E(B-V)_{\rm tot} = 0.137$\,mag is adopted for SN~2016iyc. Figure~\ref{fig:color_curve} shows the comparison of total extinction corrected $(B-V)_{0}$ colour of SN~2016iyc with other similar SNe.

\subsection{Bolometric light curves}
\label{subsec3.3}
\begin{figure*}
\centering
    \includegraphics[height=8.0cm,width=8.5cm,angle=0]{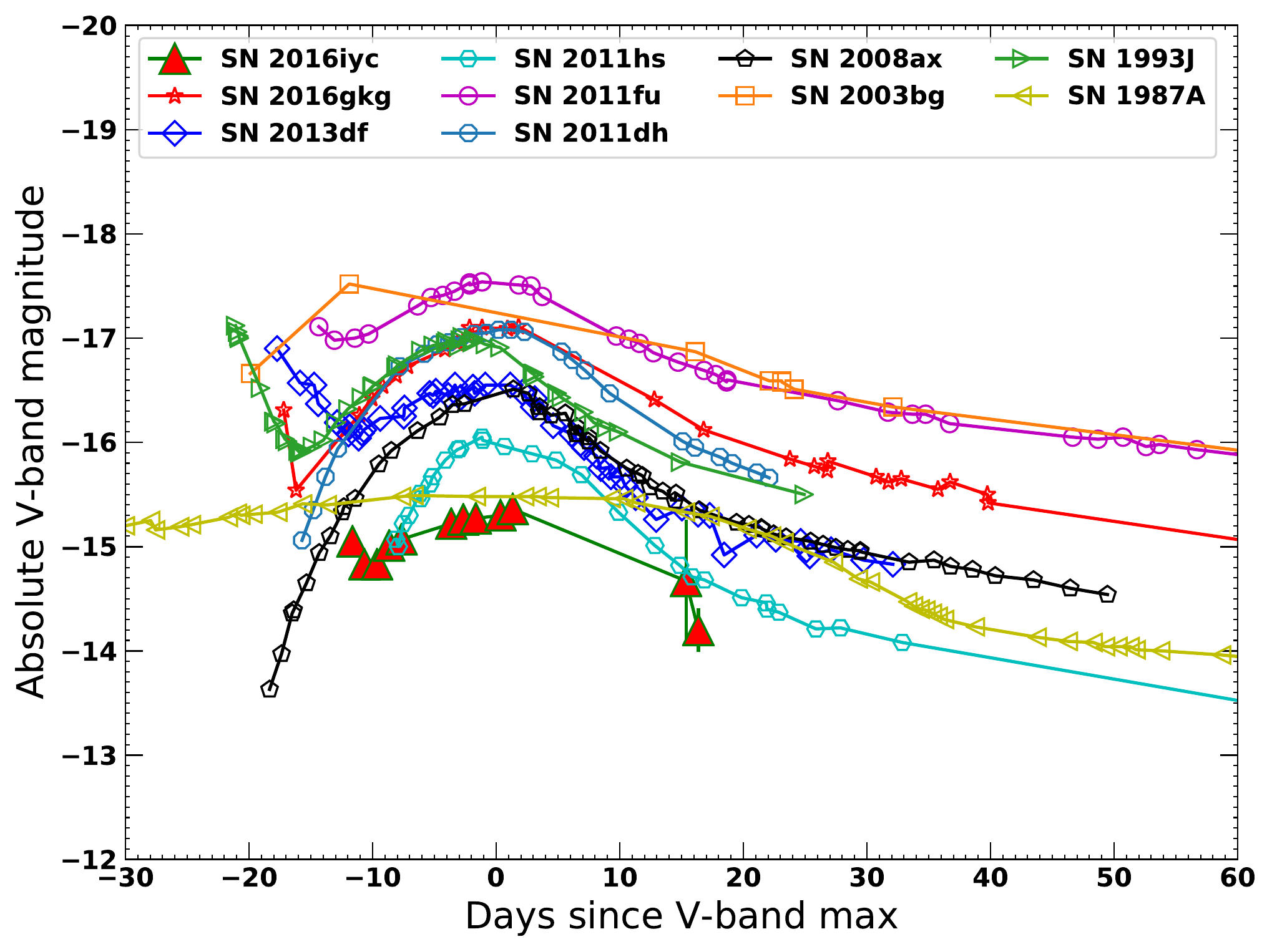}
    \includegraphics[height=8.0cm,width=8.5cm,angle=0]{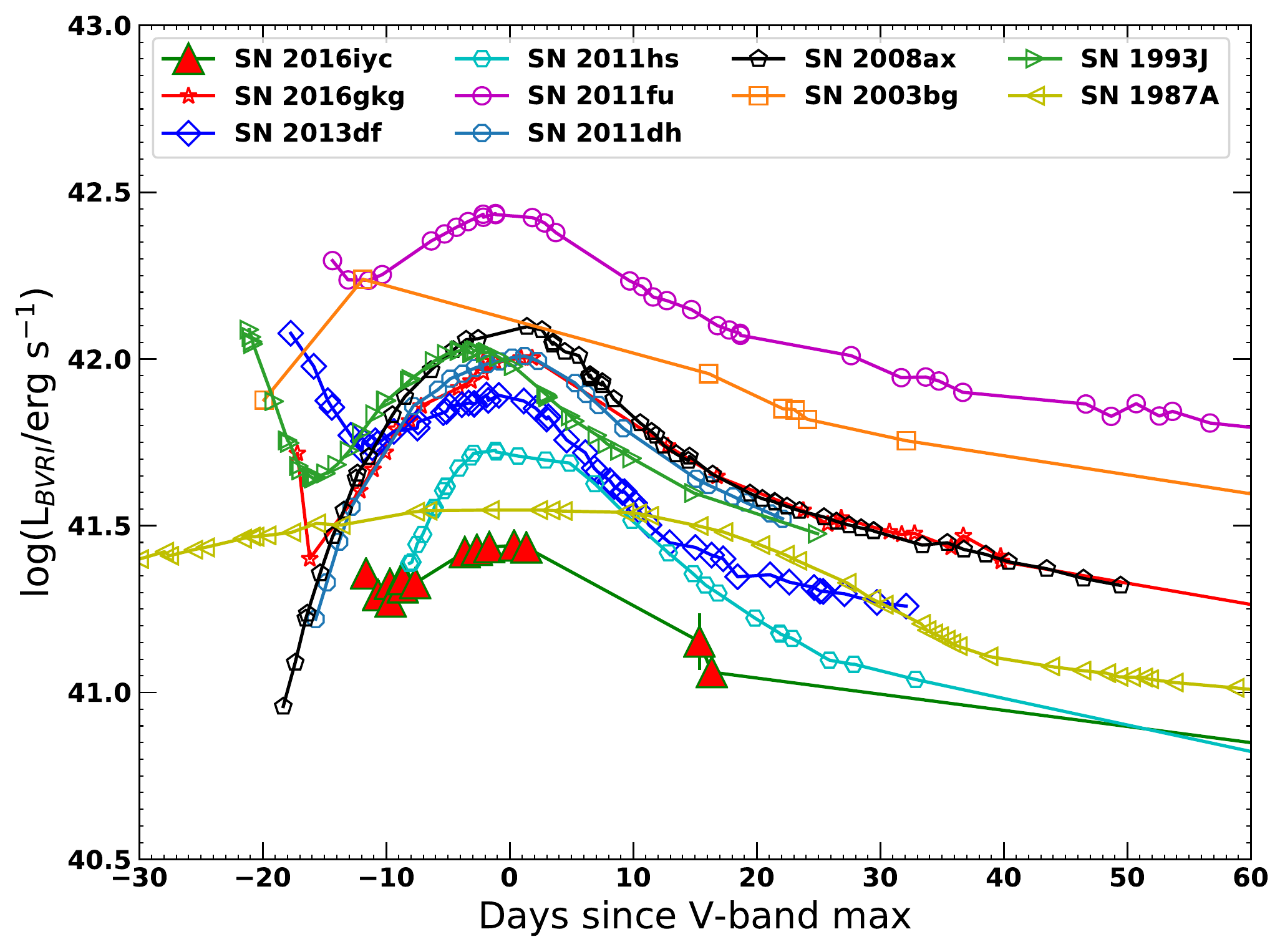}
   \caption{ {\em Left:} Comparison of the absolute $V$-band light curves of SN~2016iyc and SN~2016gkg with other well-studied SNe~IIb ( the peculiar Type II SN~1987A, included in the sample because of its low luminosity). {\em Right:} Comparison of the quasibolometric light curves of SN~2016iyc and SN~2016gkg with other well-studied SNe~IIb obtained by integrating the fluxes over the $BVRI$ bands. SN~2016iyc lies toward the faint end of SNe~IIb. The total extinction correction and distance moduli for all the SNe in the comparison sample have been taken into account while calculating these light curves.}
    \label{fig:abs_bol}
\end{figure*}

Before computing the bolometric light curves, the absolute $V$-band light curve of SN~2016iyc is compared with a few other similar SNe~IIb. The left panel of Figure~\ref{fig:abs_bol} shows that SN~2016iyc lies toward the fainter end of the distribution.

Furthermore, to obtain the quasibolometric light curve, we make use of the {\tt SUPERBOL} code \citep{Nicholl2018}. The extinction-corrected $B$, $V$, $R$, and $I$ data are provided as input to {\tt SUPERBOL}. The light curve in each filter is then mapped to a common set of times through the processes of interpolation and extrapolation. Thereafter, {\tt SUPERBOL} fits black-body curves to the SED at each epoch,  up to the observed wavelength range (4000--9000\,\si{\angstrom}), to give the quasibolometric light curve by performing trapezoidal integration.

The right-hand panel of Figure~\ref{fig:abs_bol} shows the comparison of the quasibolometric light curve of SN~2016iyc with  other well-studied SNe~IIb as listed in Table~\ref{tab:comparison_Sample}. The peak quasibolometric luminosity (log\,$(L_{BVRI})_p$) of each SN has also been calculated by fitting a third-order polynomial to the quasibolometric light curve. As indicated by the right-hand panel of Figure~\ref{fig:abs_bol}, SN~2016iyc lies toward the fainter limit of SNe~IIb in the comparison sample. It is also worth mentioning that the low-luminosity SNe with low $^{56}$Ni yields are thought to arise from progenitors having masses near the threshold mass for producing a CCSN \citep[][]{Smartt2009b}.

\begin{figure}
\includegraphics[width=\columnwidth]{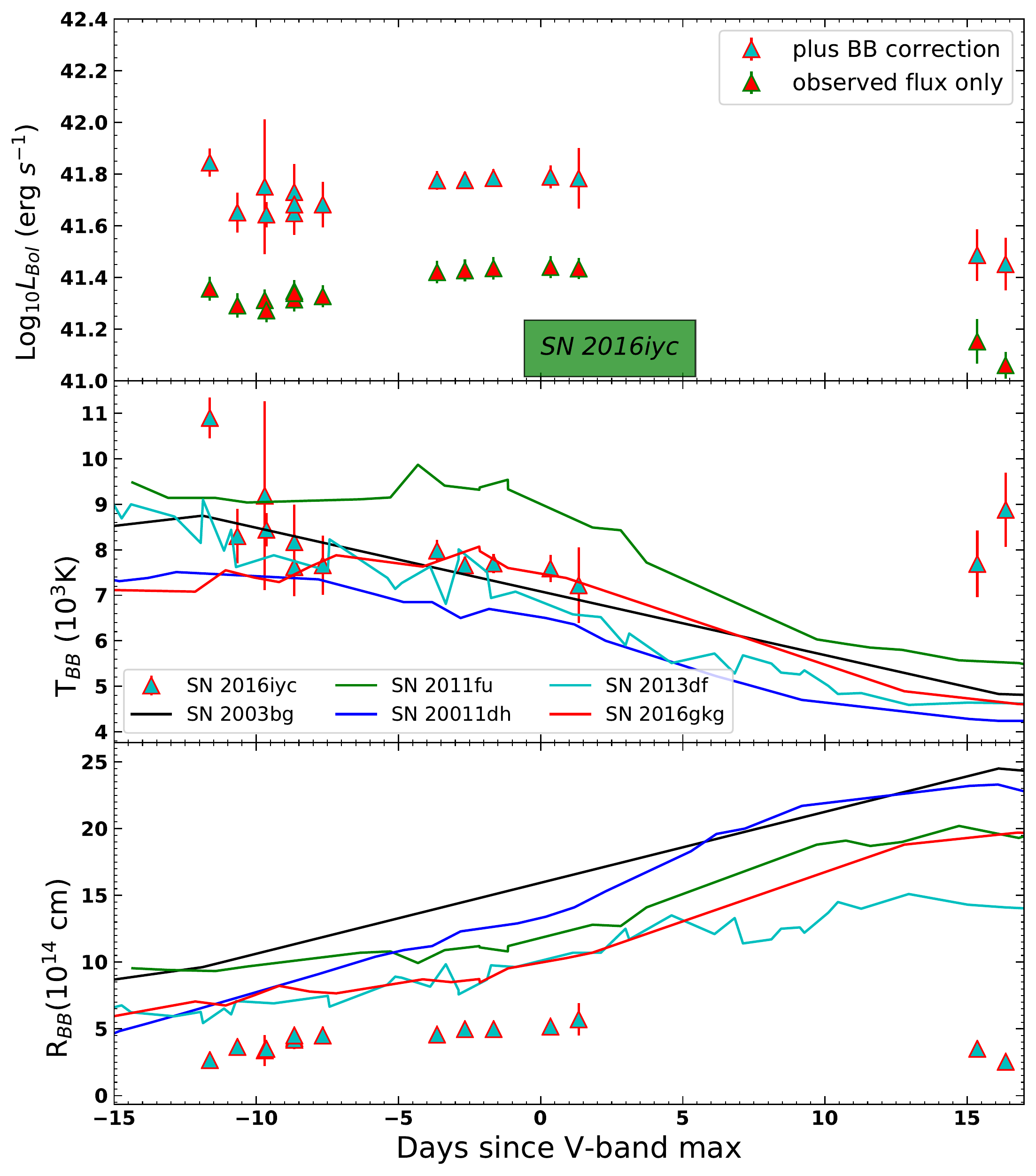}
\caption{The top panel shows the bolometric and quasibolometric light curves of SN~2016iyc. The luminosity corresponding to the late upper limits on $BVRI$ are shown later. The second and third panels from the top respectively display the black-body temperature and radius evolutions of SN~2016iyc.}
\label{fig:BB_param_SN2016iyc}
\end{figure}

Furthermore, the bolometric luminosity light curve of SN~2016iyc is also produced after considering the additional black-body corrections to the observed $BVRI$ quasibolometric light curve, by fitting a single black body to observed fluxes at a particular epoch and integrating the fluxes trapezoidally for a wavelength range of 100--25,000\,\AA\ using {\tt SUPERBOL}. Figure \ref{fig:bb_fits} shows the black-body fits to the SED of SN~2016iyc. The top panel of Figure~\ref{fig:BB_param_SN2016iyc} shows the resulting quasibolometric and bolometric light curves of SN~2016iyc.

\subsection{Temperature and radius evolution}

From {\tt SUPERBOL}, the black-body temperature ($T_{\rm BB}$) and radius ($R_{\rm BB}$) evolution of SN~2016iyc are also obtained. During the initial phases, the photospheric temperature is high, reaching $\sim 10,900$\,K at $-10.63$\,d. Furthermore, the SN seems to evolve very rapidly; its temperature quickly drops to $\sim 7600$\,K in only a few days around $-7.7$\,d and then remains nearly constant (Figure~\ref{fig:BB_param_SN2016iyc}, second panel from top). Along with SN~2016iyc, the temperature evolution of a few more similar SNe~IIb  are also shown in this panel. The black-body temperature of SN~2016iyc seems to follow the typical temperature evolution as seen in SNe~IIb.  

A conventional evolution in radius is also seen. Initially, at an epoch of $-10.63$\,d, the black-body radius is $2.64 \times 10^{14}$\,cm. Thereafter, the SN expands and its radius increases, reaching a maximum radius of $\sim$\,$5.8 \times 10^{14}$\,cm, beyond which the photosphere seems to recede into the SN ejecta (Figure~\ref{fig:BB_param_SN2016iyc}, third panel from top). Along with SN~2016iyc, the black-body radius evolution of a few more similar SNe~IIb are also shown. SN~2016iyc seems to exhibit anomalous behaviour, with its black-body radii at various epochs being the smallest among other similar SNe~IIb. This result can be attributed to the low ejecta velocity of SN~2016iyc.

\section{Spectral studies of SN~2016iyc}
\label{sec:Spectral}

In this section, we identify the signatures of various lines by modelling the only available spectrum of SN~2016iyc using {\tt SYN++} \citep[][]{Branch2007, Thomas2011}. We discuss various spectral features of SN~2016iyc, and the spectrum is also compared with other similar SNe.

\subsection{Spectral modelling}
\begin{figure}
\centering
\includegraphics[width=\columnwidth]{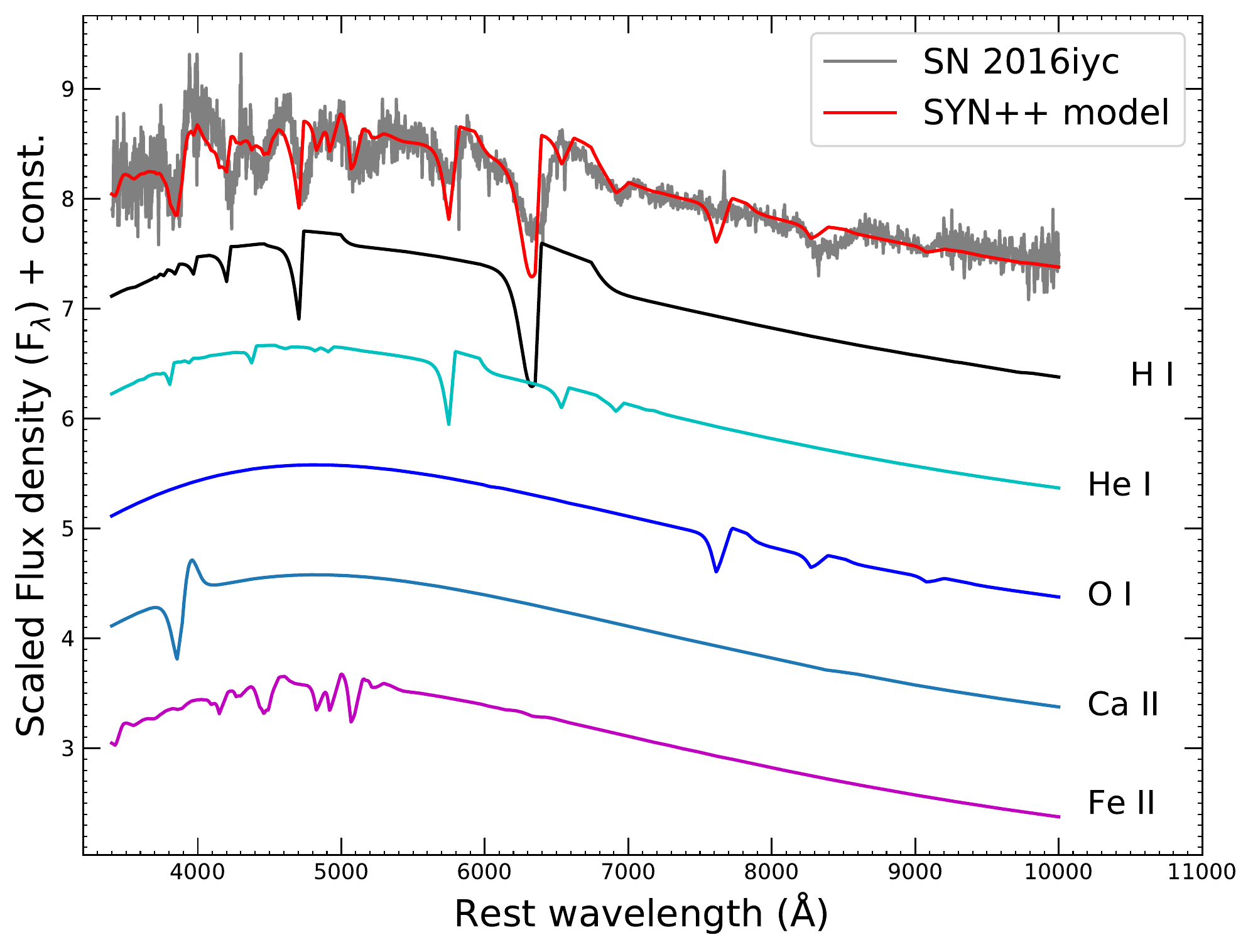}
\caption{{\tt SYN++} modelling of the spectrum of SN~2016iyc at a phase of $-6.6$\,d. The effects of various elements present in the SN ejecta and contributing to the spectrum are also displayed individually.}
\label{fig:syn++}
\end{figure}

A single optical spectrum of SN~2016iyc was obtained on 2016 Dec. 23 with the Kast double spectrograph \citep[][]{MillerStone1993} mounted on the 3\,m Shane telescope at Lick Observatory. Figure~\ref{fig:syn++} shows the spectral modelling of it, corresponding to a phase of $-6.6$\,d. The individual lines corresponding to various elements and ions are also indicated for better identification of the features. The profiles of H$\alpha$ near 6563\,\AA, He\,I near 5876\,\AA, and Ca\,II~H\&K are very nicely reproduced by {\tt SYN++} modelling. A very strong H$\alpha$ feature near 6563\,\AA\ classifies SN~2016iyc as an SN~IIb. The observed velocities obtained using H$\alpha$, He\,I, and Fe\,II features in the spectrum are $\sim 10,000$\,km\,s$^{-1}$, $\sim 6700$\,km\,s$^{-1}$, and $\sim 6100$\,km\,s$^{-1}$ (respectively), while the respective velocities of these lines from {\tt SYN++} modelling are 10,100\,km\,s$^{-1}$, 6800\,km\,s$^{-1}$, and 6100\,km\,s$^{-1}$, very close to the observed ones. The parameterisation velocity and photospheric velocity used to produce the {\tt SYN++} model are 6000\,km\,s$^{-1}$ and 6100\,km\,s$^{-1}$, respectively. Also, a photospheric temperature of 6300\,K is employed to produce the model spectrum.

\begin{figure}
\includegraphics[width=\columnwidth]{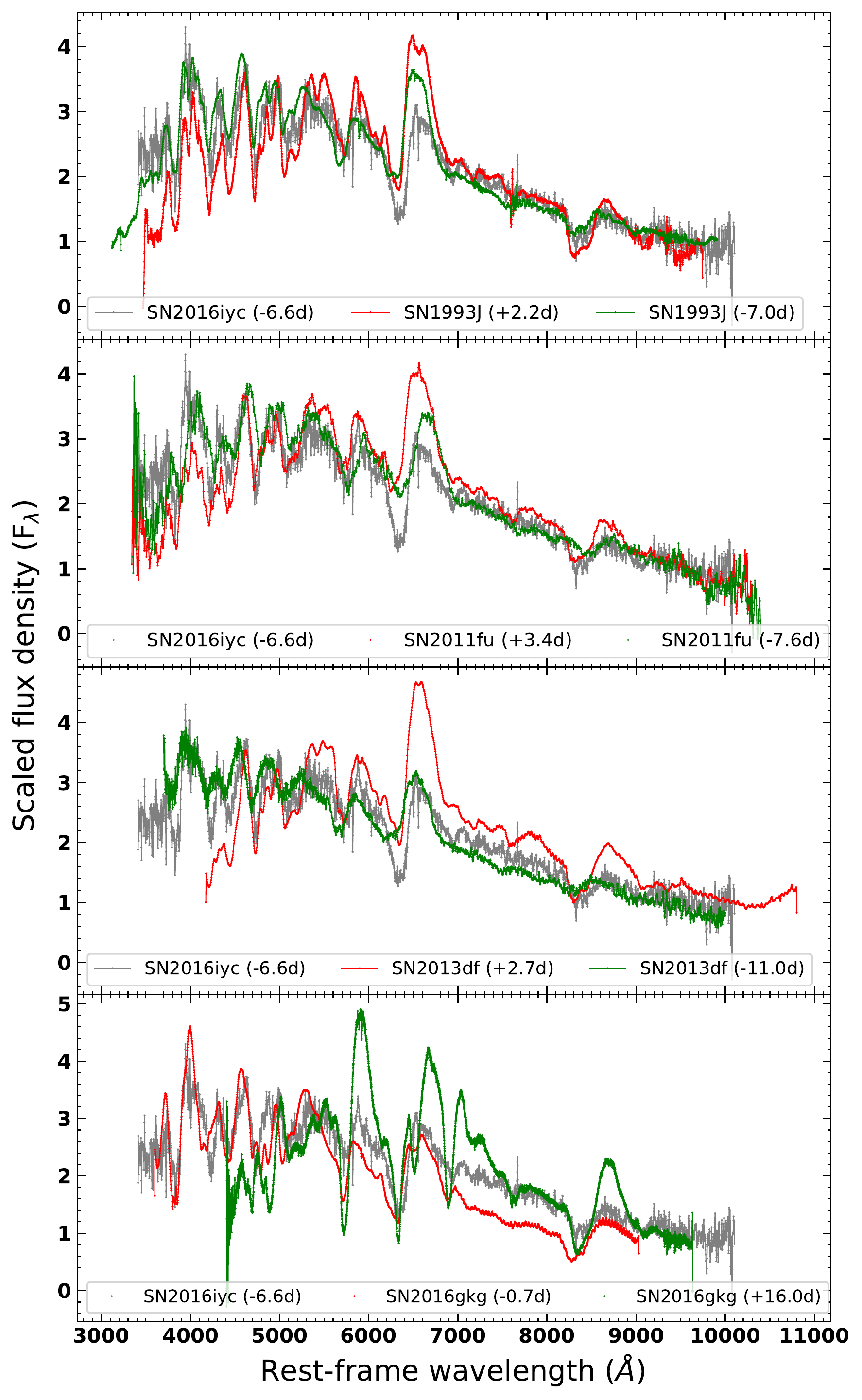}
\caption{Comparison of the $-6.6$\,d spectrum of SN~2016iyc with spectra of other well-studied SNe~IIb, including SN~1993J, SN~2011fu, SN~2013df, and SN~2016gkg. }
\label{fig:spec_comparison}
\end{figure}

\subsection{Spectral comparison}
Figure~\ref{fig:spec_comparison} shows a comparison of the normalised spectrum of SN~2016iyc with other well-studied SNe~IIb. The top plot displays the comparison with the spectra of SN~1993J at +2.3\,d and $-7.0$\,d; we see that the spectral features of SN~2016iyc closely resemble those of SN~1993J spectra. In the second panel from the top, the spectrum of SN~2016iyc is compared with spectra of SN~2011fu at +3.4\,d and $-7.0$\,d; the match is close, except for the H$\alpha$ feature where the spectra of SN~2011fu are slightly off. The third panel from the top shows the spectral comparison of SN~2016iyc with spectra of SN~2013df at epochs of +2.7\,d and $-11.0$\,d, revealing a good match with the $-11.0$\,d spectrum. The progenitor of SN~2013df is also thought to be arising from the lower-mass end. In the bottom panel, the spectrum of SN~2016iyc is compared with spectra of SN~2016gkg at epochs of $-0.7$\,d and +16\,d. The $-0.7$\,d spectrum of SN~2016gkg resembles the spectrum of SN~2016iyc toward the bluer side, while features in the redder part of the spectrum are slightly off. The +16\,d spectrum of SN~2016gkg does not show a good resemblance with the spectrum of SN~2016iyc. From Figure~\ref{fig:spec_comparison}, we conclude that the spectrum of SN~2016iyc shows close resemblance with the spectra of other well-studied SNe~IIb, thereby providing good evidence for the classification of SN~2016iyc as an SN~IIb.

\section{Possible progenitor modelling and the results of synthetic explosions for SN~2016iyc}
\label{sec:mesa_snec}
\begin{figure*}
\centering
    \includegraphics[height=8.0cm,width=8.5cm,angle=0]{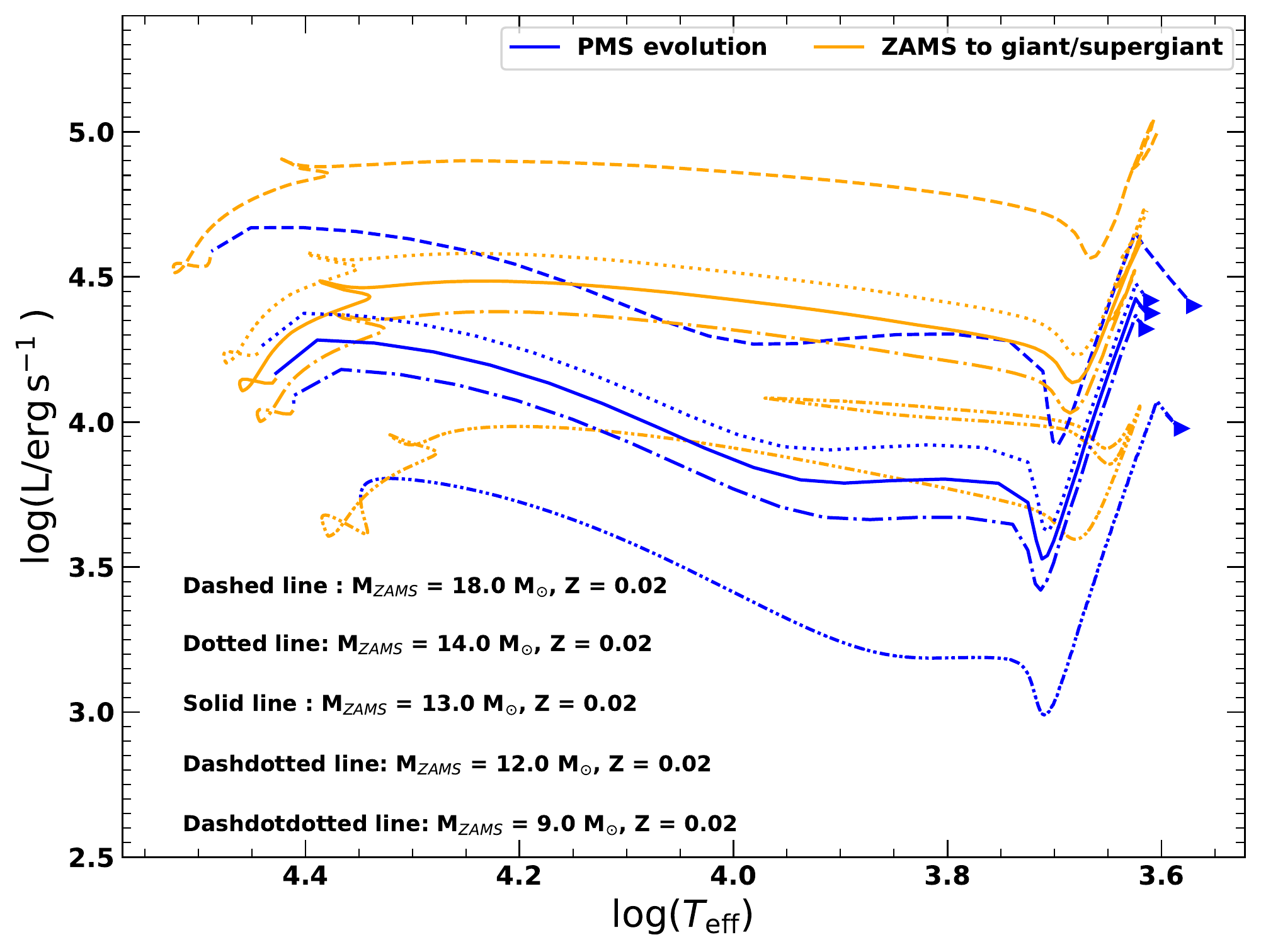}
    \includegraphics[height=8.0cm,width=8.5cm,angle=0]{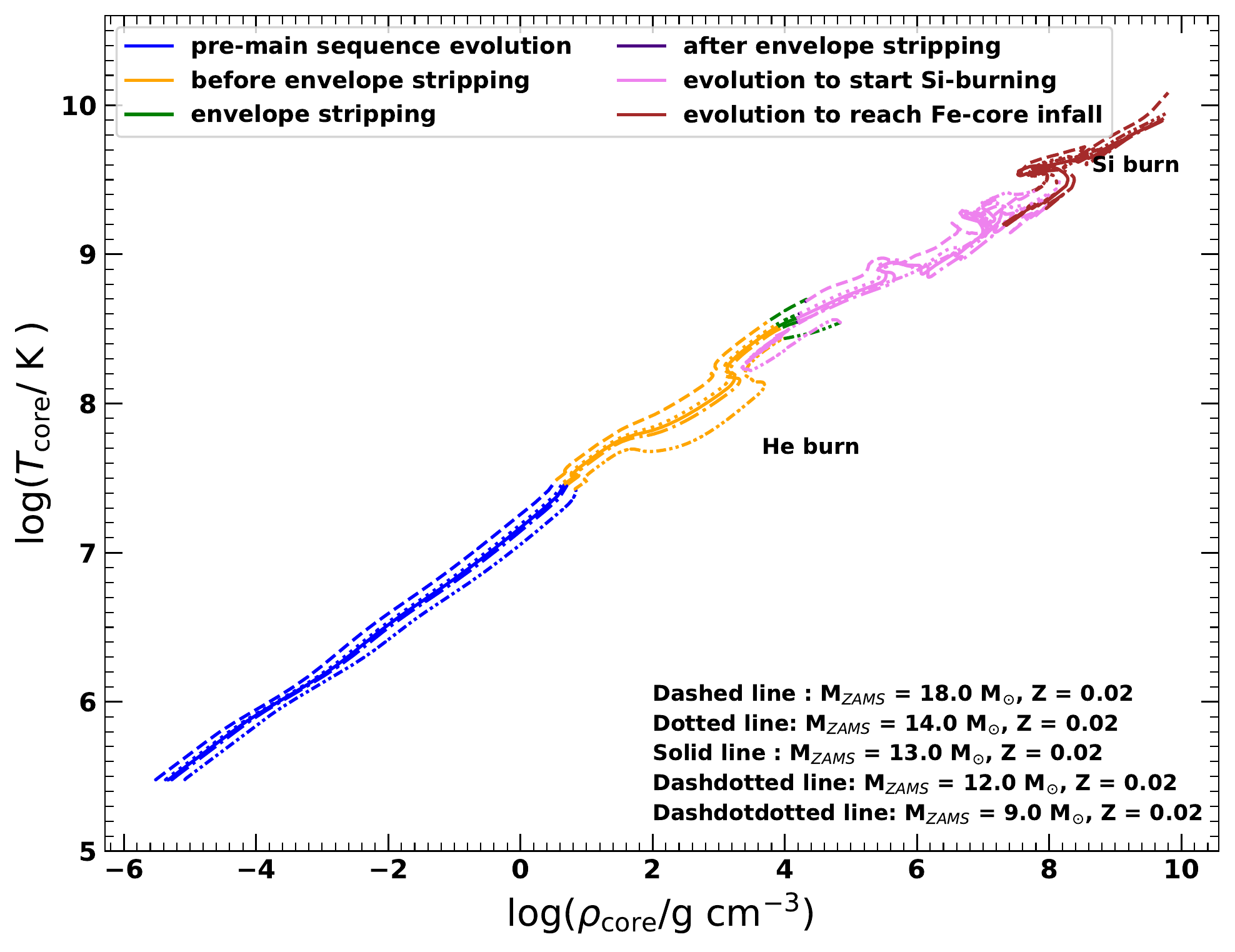}
   \caption{ {\em Left:} The evolution of 9.0\,M$_{\odot}$, 12\,M$_{\odot}$, 13\,M$_{\odot}$, 14\,M$_{\odot}$, and 18.0\,M$_{\odot}$ ZAMS progenitors with $Z = 0.02$ on the HR diagram. The models begin evolution on the pre-main sequence (blue curve), then reach the main sequence and evolve until they become giants/supergiants with being ready to strip their outer envelopes (orange curve). {\em Right:} The variation of core temperature with core density as the models evolve through various phases on the HR diagram. The core-He and core-Si burning phases are marked. The onset of Fe-core infall in the models is marked by the core temperatures and core densities reaching above $10^{10}$\,K and $10^{10}$\,g\,cm$^{-3}$, respectively.}
    \label{fig:HR_Rho_T}
\end{figure*}

To constrain the physical properties of the possible progenitor of SN~2016iyc, we attempted several progenitor models. Following the available literature, SN~2016iyc lies near the faint limit (see Table~\ref{tab:comparison_Sample}), with $M_{\rm ej}$ also close to the lowest limit (Table~\ref{tab:mejecta_comparison}). As mentioned earlier, low-luminosity SNe with low $^{56}$Ni production are thought to arise from progenitors having masses near the threshold mass for producing CCSNe \citep[][]{Smartt2009}. With low $M_{\rm ej}$ among typical SNe~IIb and having intrinsically low luminosity, we started with the nearly lowest possible ZAMS progenitor mass of 9\,M$_{\odot}$ for a Type IIb SN. Starting from the pre-main sequence, the model could be evolved up to the onset of core collapse. But the 9\,M$_{\odot}$ model at pre-SN phase in our simulation is very compact, having a radius of only 0.14\,R$_{\odot}$. Such a compact progenitor cannot generate the generic extended-SBO feature of typical SNe~IIb. Furthermore, no direct observational signatures have been found for an SN~IIb arising from a progenitor having ZAMS mass $\leq 11$\,M$_{\odot}$, so we do not make any further attempt to model progenitors having masses $\leq 11$\,M$_{\odot}$. Thus, we select models having ZAMS masses of 12, 13, and 14\,M$_{\odot}$, and evolve them up to the onset of core collapse. Such models originating from the lower limits of progenitor mass systems lack sufficiently strong winds to suffer much stripping; thus, the models are artificially stripped to mimic the effect of a binary companion. A brief description of our models is provided below.

\begin{table}
\caption {Ejecta masses of various SNe~IIb and SN~2016iyc.}
\label{tab:mejecta_comparison}
{\scriptsize
\begin{tabular}{ccccccccccccc}
\hline\hline
	SN name    &	$M_{\rm ej}$	&	source  \\
\hline

SN~1993J  	 	&	1.9--3.5 	& \citet[][]{Young1995} \\\\

SN~2003bg  	 	&	4	        & \citet[][]{Mazzali2009} \\\\

SN~2008ax  	    &	2--5	        & \citet[][]{Taubenberger2011}   \\\\

SN~2011dh  	 	&	1.8--2.5  	& \citet[][]{Bersten2012} \\\\

SN~2011fu 	 	&	3.5      	& \citet[][]{Morales2015} \\\\

SN~2011hs  	 	&	1.8     	& \citet[][]{Bufano2014}   \\\\

SN~2013df  	 	&	0.8--1.4 	& \citet[][]{Morales2014} \\\\

SN~2016gkg 	    &	3.4     	& \citet[][]{Bersten2018}	 \\\\

SN~2016iyc 	    &	1.2     	& \citet[][]{Zheng2022}   \\

\hline\hline
\end{tabular}}
\end{table}

We first evolve the nonrotating 9, 12, 13, and 14\,$M_{\odot}$ ZAMS stars until the onset of core collapse, using the one-dimensional stellar evolution code {\tt MESA}\citep[][] {Paxton2011,Paxton2013,Paxton2015,Paxton2018}.

For the 9\,$M_{\odot}$ model, $\alpha_{\rm MLT} = 2.0$ is used throughout the evolution, except for the phase when the model evolves to reach the beginning of core-Si burning (i.e., in the {\tt inlist\_to\_si\_burn} file), where $\alpha_{\rm MLT} = 0.01$ is used. At this phase, the evolution of the models is extremely sensitive to this $\alpha_{\rm MLT}$, since even a slight change (say, 0.02) results in the failure of the beginning of core-Si burning. Although $\alpha_{\rm MLT} = 0.01$ seems to be very low, this is required for the successful evolution of considered models through the last phases of their evolution. Furthermore, as mentioned by \citet[][]{Joyce2018}, $\alpha_{\rm MLT}$ is neither a physical constant nor a computational one; it is rather a free parameter, so its value must be determined on an individual basis in each stellar evolution code. Thus, as $\alpha_{\rm MLT} = 0.01$ is helpful for evolving the models beyond the beginning of core-Si burning, it is acceptable.
For the 12, 13, and 14\,$M_{\odot}$ models, $\alpha_{\rm MLT} = 3.0$ is used throughout the evolution.

Convection is modelled using the mixing theory of \citet[][]{Henyey1965}, adopting the Ledoux criterion. Semiconvection is modelled following \citet[][]{Langer1985} with an efficiency parameter of $\alpha_{\mathrm{sc}} = 0.01$. For the thermohaline mixing, we follow \citet[][]{Kippenhahn1980}, and set the efficiency parameter as $\alpha_{\mathrm{th}} = 2.0$. We model the convective overshoot with the diffusive approach of \citet[][]{Herwig2000}, with $f= 0.001$ and $f_0 = 0.007$ for all the convective cores and shells. We use the \say{Dutch} scheme for the stellar wind, with a scaling factor of 1.0. The \say{Dutch} wind scheme in MESA combines results from several papers. Specifically, when $T_{\mathrm{eff}} > 10^4$\,K and the surface mass fraction of H is greater than 0.4, the results of \citet[][]{Vink2001} are used, and when $T_{\mathrm{eff}} > 10^4$\,K and the surface mass fraction of H is less than 0.4, the results of \citet[][]{Nugis2000} are used. In the case when $T_{\mathrm{eff}} < 10^4$\,K, the \citet[][]{dejager1988} wind scheme is used.

\begin{figure*}
\centering
    \includegraphics[height=8.0cm,width=8.5cm,angle=0]{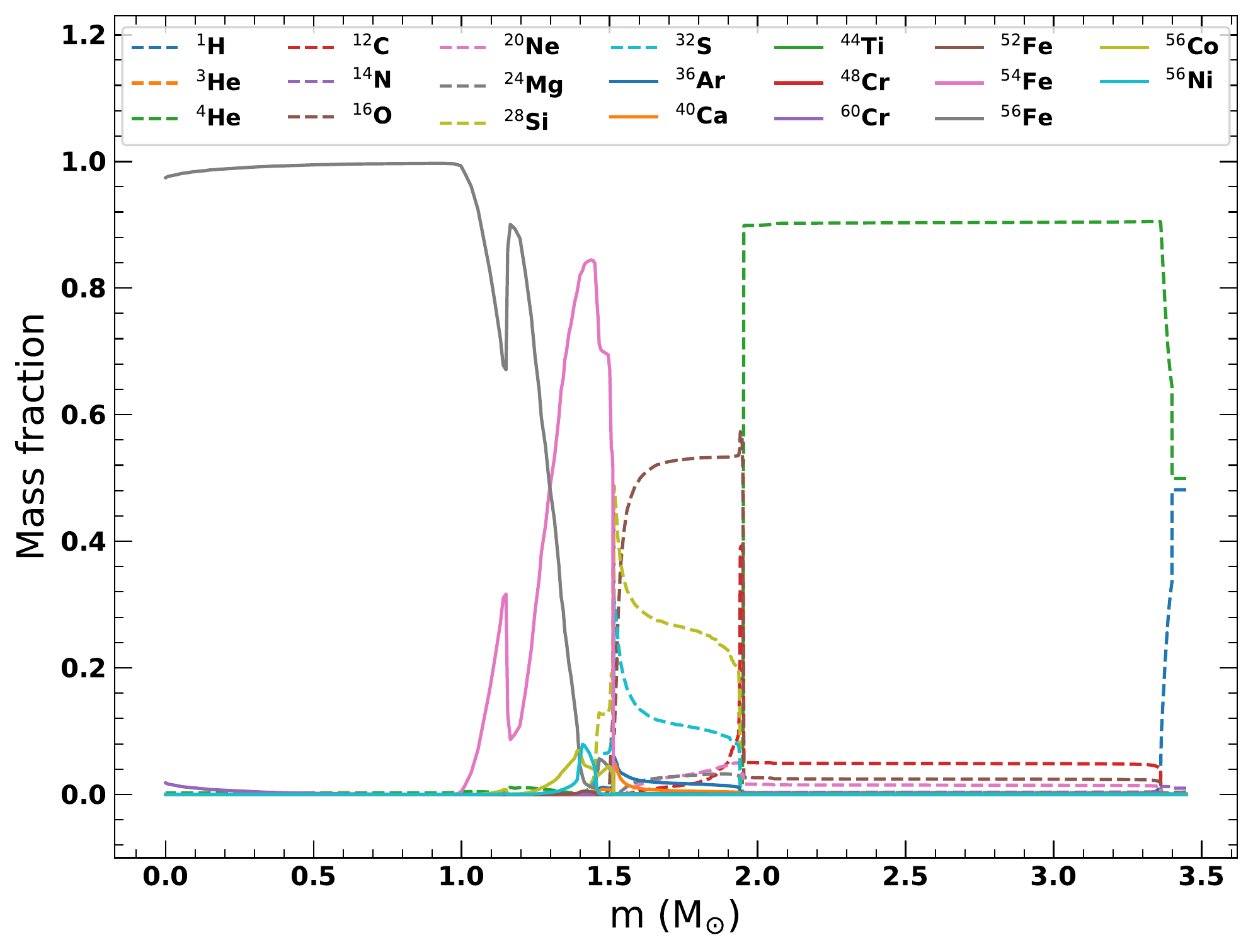}
    \includegraphics[height=8.0cm,width=8.5cm,angle=0]{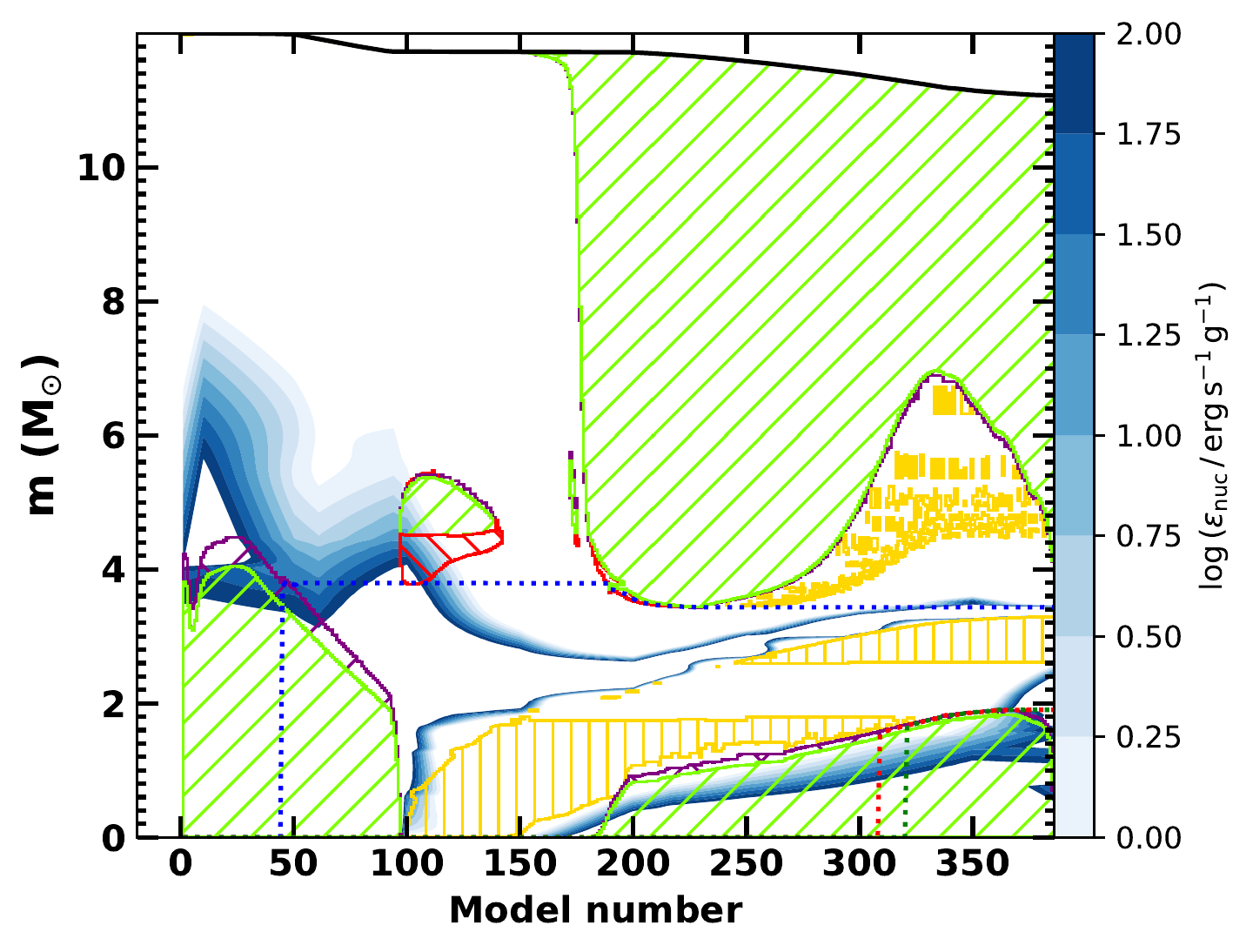}
   \caption {{\em Left:} The mass fractions of a few key elements when the 12\,M$_{\odot}$ ZAMS progenitor model with $Z = 0.02$ has just reached the stage of Fe-core infall. Notice the very high mass fraction of $^{56}$Fe in the core compared to other species. {\em Right:}  The Kippenhahn diagram of the same model for a period from the beginning of main-sequence evolution to the stage when the model is ready to be stripped.}
    \label{fig:mass_Kipp}
\end{figure*}

\begin{figure*}
\centering
    \includegraphics[height=8.0cm,width=8.5cm,angle=0]{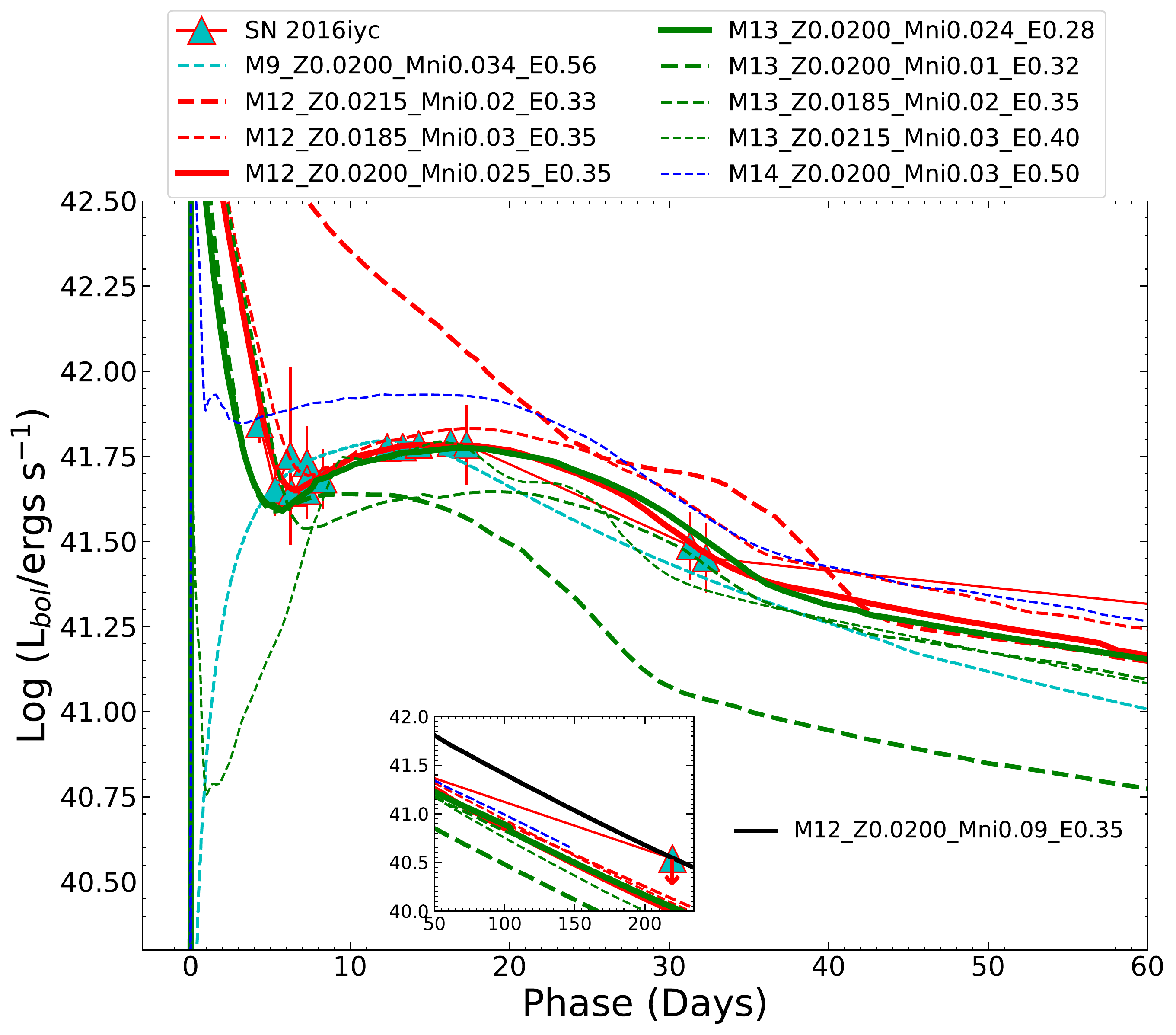}
    \includegraphics[height=8.0cm,width=8.5cm,angle=0]{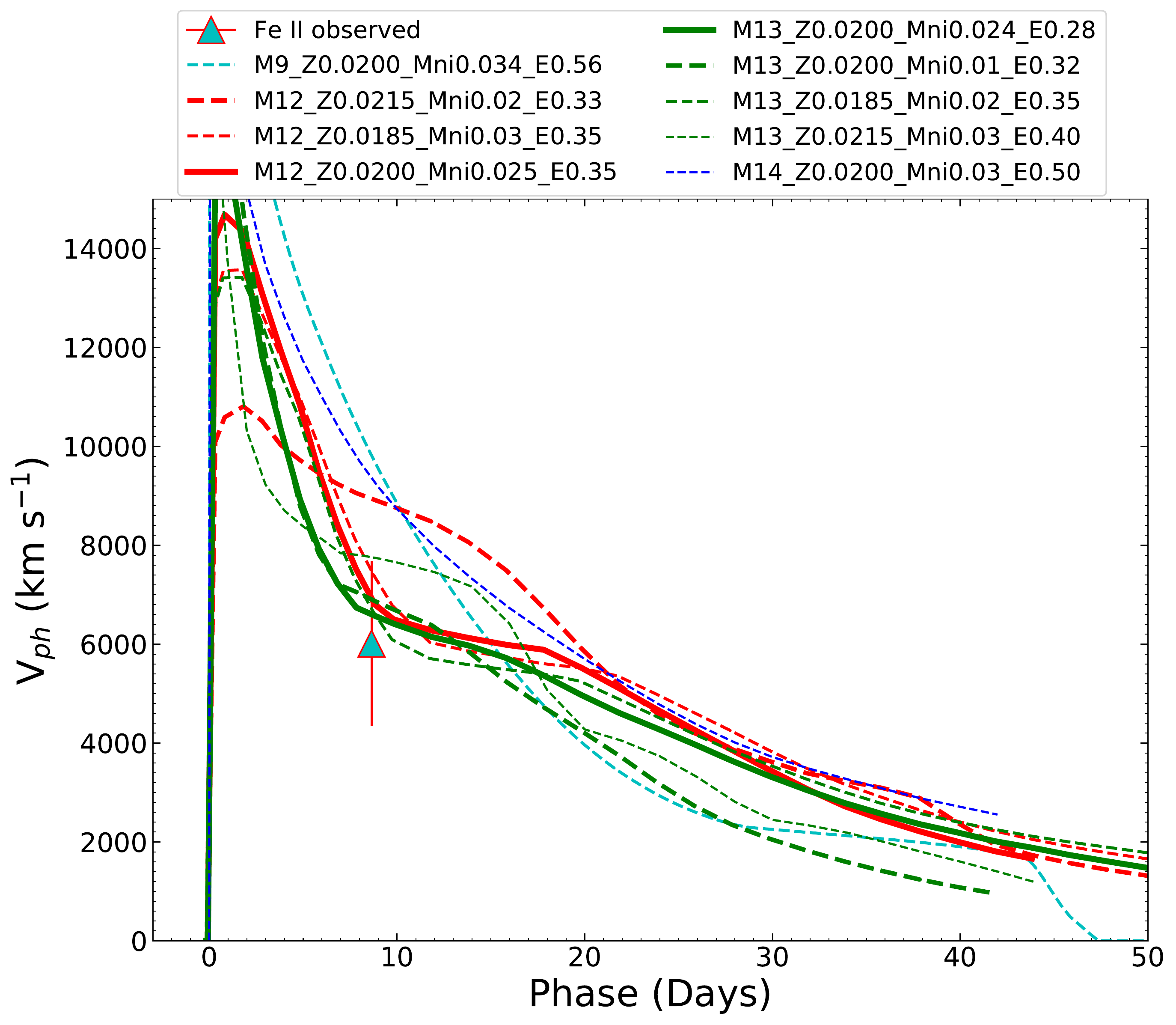}
   \caption {The results of the synthetic explosions produced using {\tt STELLA/SNEC} by assuming  9\,M$_{\odot}$, 12\,M$_{\odot}$, 13\,M$_{\odot}$, and  14\,M$_{\odot}$ ZAMS stars as the possible progenitors for SN~2016iyc. {\em Left:} The bolometric luminosity light curves corresponding to different models having different metallicities, explosion energies, and nickel masses compared with the observed bolometric light curve of SN~2016iyc. {\em Right:} The corresponding photospheric velocity evolution comparison. In both the panels, the ``Phase'' is (approximate) days since the explosion. Following \citet[][]{Zheng2022}, the adopted phase of explosion is $-4.64^{+0.67}_{-0.76}$ since first detection that corresponds to MJD $57736.47^{+0.67}_{-0.76}$. The velocities produced by the models are well within the error bar of the observed photospheric velocities indicated by the Fe\,II line velocity.}
    \label{fig:hydrodynamical_lc}
\end{figure*}

SNe~IIb have been considered to originate from massive stars which have retained a significant amount of their hydrogen envelope. We have adopted a mass-loss rate of $\dot{M} \gtrsim 10^{-4}$\,M$_{\odot}\,\mathrm{yr}^{-1}$ to artificially strip the models until the final $M_{\rm H}$ reaches in range of (0.013--0.055)\,M$_{\odot}$. Such extensive mass loss rates are supported by the studies performed by \citet[][]{Ouchi2017}, where they mention that the binary scenario for the progenitors of SNe~IIb leads to such high mass-loss rates. Furthermore, \citet[][]{VanLoon2005}  have also reported extensive mass-loss rates reaching $\dot{M} = 10^{-4}$\,M$_{\odot}\,\mathrm{yr}^{-1}$, solely by a single stellar wind.

Once the models have stripped off upto the specified limit of the H envelope, we switch off the artificial mass loss and further evolve the models until the onset of core collapse. Corresponding to various ZAMS masses, the amount of remaining H varies. Massive progenitors with a similar rate of stripping as less-massive progenitors will retain a larger amount of H. In our simulations, the specified limit on H mass depends primarily on (a) the model's ability to evolve up to the stage of core collapse by retaining the specified amount of H (also, there is a limit on stripping), and (b) the radius of the pre-SN progenitor. If we need a compact progenitor, the amount of retained H is less, and if pre-SN progenitors are required to have extended envelopes, the amount of retained H is more.

The evolution of the models using {\tt MESA} takes place in various steps. The models start to evolve on the pre-main sequence and reach the main sequence. The arrival of the models on the main sequence is marked when the ratio of the luminosity due to nuclear reactions and the total luminosity of the models is 0.8. Later, the models further evolve on the main sequence, becoming giants or supergiants. As a next step, artificial stripping of the models is performed, after which they are allowed to settle down. Once the stripping of the models reaches the specified H-envelope mass limit and the models have settled down, they further evolve until the ignition of Si~burning in their core. Once the Si~burning has started in the core, the models begin to develop inert iron ($^{56}$Fe) cores responsible for their cores to collapse.

The evolution of such models having ZAMS masses of 9, 12, 13, and 14\,M$_{\odot}$ with metallicity $Z = 0.0200$ on the Hertzsprung-Russell (HR) diagram is illustrated in the left panel of Figure~\ref{fig:HR_Rho_T}. We simulated a total of 9 models covering progenitor masses of 9--14\,M$_{\odot}$ and also covering subsolar to supersolar metallicity wherever necessary. The pre-explosion properties using {\tt MESA} and explosion properties using {\tt STELLA/SNEC} are listed in Table~\ref{tab:MESA_MODELS}. The models have been so named that they include informations of ZAMS mass, metallicity, $M_{\rm Ni}$, and $E_{\rm exp}$. Thus, the model {\tt M9.0\_Z0.0200\_Mni0.034\_E0.56} has a ZAMS mass of 9\,M$_{\odot}$, $Z = 0.0200$, $M_{\rm Ni} = 0.034$\,M$_{\odot}$, and $E_{\rm exp} = 0.56\times10^{51}$\,erg.

The right-hand panel of Figure~\ref{fig:HR_Rho_T} shows the variation of core temperature ($T_{\rm core}$) with core density ($\rho_{\rm core}$) as the models evolve through various phases on the HR diagram. The core-He and core-Si burning phases are marked. The onset of core collapse is marked by $T_{\rm core}$ and $\rho_{\rm core}$ reaching above $\sim10^{10}$\,K and $10^{10}$\,g\,cm$^{-3}$, respectively. The left panel of Figure~\ref{fig:mass_Kipp} shows the mass fractions of various species present when the 12\,M$_{\odot}$ model (with $Z = 0.0200$) has achieved Fe-core infall. The core is composed mainly of $^{56}$Fe with negligible fractions of other species. Significant fractions of heavier metals are seen toward the surface of the pre-explosion progenitor. The right-hand panel of Figure~\ref{fig:mass_Kipp} shows the Kippenhahn diagram for the 12\,M$_{\odot}$ model (with $Z = 0.0200$) for a period from the beginning of main-sequence evolution to the stage when the model is ready to begin envelope stripping. In this figure, the convective regions are marked by the hatchings with the logarithm of the specific nuclear energy generation rate ($\epsilon_{\rm nuc}$) inside the stellar interiors marked with the blue colours. The dark-yellow regions indicate the stellar interior where the thermohaline mixing is going on.

\begin{table*}
\caption{{\tt MESA} model and {\tt STELLA/SNEC} explosion parameters of various models for SN~2016iyc.}
\label{tab:MESA_MODELS}
\begin{center}
{\scriptsize
\begin{tabular}{ccccccccccccc} 
\hline\hline
Model Name	& $M_{\rm ZAMS}$	& $Z$  & $M_{\mathrm{H}}^{a}$ & $R_{\mathrm{0}}^{b}$ & $f_{ov}^{c}$ &	$M_{\mathrm{f}}^{d}$	& $M_\mathrm{ci}^{e}$	&	$M_\mathrm{cf}^{f}$ & $M_{\mathrm{ej}}^{g}$	&	$M_{\mathrm{Ni}}^{h}$ &	$E_{\mathrm{exp}}^{i}$ 	\\
	&	(M$_{\odot}$) &	&	(M$_{\odot}$)	 &  (R$_{\odot}$) &  & (M$_{\odot}$)	&	(M$_{\odot}$) 	&	(M$_{\odot}$)  & (M$_{\odot}$) & (M$_{\odot}$) & 	($10^{51}$\,erg) 	\\
\hline
\hline

M9.0\_Z0.0200\_Mni0.034\_E0.56     &	9.0  	&	0.0200  &  0.013   & 0.14  & 0.007  &    2.17  & 1.4 & 1.4 &  0.77 & 0.034 &  0.56		\\

M12.0\_Z0.0215\_Mni0.02\_E0.33     &	12.0  	&	0.0215  &  0.035   & 596  & 0.007  &    3.96  & 1.54 & 1.54 & 2.42 & 0.02 &  0.33		\\

M12.0\_Z0.0185\_Mni0.03\_E0.35     &	12.0  	&	0.0185  &  0.055   & 315 & 0.007  &    3.49  & 1.46 & 1.46 & 2.03 & 0.03 &  0.35		\\

M12.0\_Z0.0200\_Mni0.025\_E0.35     &	12.0  	&	0.0200  &  0.05    & 300 & 0.007  &    3.45  & 1.52 &  1.52 & 1.93 & 0.025 &  0.35		\\

M12.0\_Z0.0200\_Mni0.09\_E0.35     &	12.0  	&	0.0200  &  0.05    & 300 & 0.007  &    3.45  & 1.52 & 1.52 & 1.93 & 0.09 &  0.35		\\

M13.0\_Z0.0200\_Mni0.024\_E0.28     &	13.0  	&	0.0200  &  0.04    & 204 & 0.007  &    3.79  & 1.64 & 1.90 & 1.88 & 0.024 &  0.28 		\\

M13.0\_Z0.0200\_Mni0.01\_E0.32     &	13.0  	&	0.0200  &  0.04    & 204 & 0.007  &    3.79  & 1.64 & 1.64 & 2.15 & 0.01 &  0.32 		\\

M13.0\_Z0.0185\_Mni0.02\_E0.35     &	13.0  	&	0.0185  &  0.06    & 318 & 0.007  &    3.92  & 1.53 & 1.56 & 2.36 & 0.02 &  0.35		\\

M13.0\_Z0.0215\_Mni0.03\_E0.40     &	13.0  	&	0.0215  &  0.015   & 10  & 0.007  &    3.81  & 1.61 & 1.62 &  2.19 & 0.03 &  0.40		\\

M14.0\_Z0.0200\_Mni0.03\_E0.50     &	14.0  	&	0.0200  &  0.03    & 55  & 0.007  &    4.23  & 1.54 & 1.54 & 2.69 & 0.03 &  0.50		\\

\hline\hline
\end{tabular}}
\end{center}
{$^a$ Amount of hydrogen retained after stripping.
$^b$Pre-SN progenitor radius.
$^c$Overshoot parameter.
$^d$Final mass of pre-SN model.
$^e$Initial mass of the central remnant.
$^f$Final mass of the central remnant.
$^g$Ejecta mass.
$^h$Nickel mass.
$^i$Explosion energy.}\\

\end{table*} 

Using the progenitor models on the verge of core collapse obtained through {\tt MESA}, we carried out radiation hydrodynamic calculations to simulate the synthetic explosions. For this purpose, we used {\tt STELLA}\citep[][]{Blinnikov1998, Blinnikov2000, Blinnikov2006} and {\tt SNEC} \citep[][]{Morozova2015}. {\tt STELLA} solves the radiative transfer equations in the intensity momentum approximation in each frequency bin, while {\tt SNEC} is a one-dimensional Lagrangian hydrodynamic code that solves the radiation energy transport equations in the flux-limited diffusion approximation. {\tt STELLA} and {\tt SNEC}, both generate the bolometric light curve and the photospheric velocity evolution of the SN, along with a few other observed parameters. The radioactive decay of $^{56}$Ni to $^{56}$Co is considered to be one of the prominent mechanisms for powering the primary peak of SNe~IIb. Both of these codes incorporate this model by default. Thus in this section, we model the entire bolometric light curve of the SN~2016iyc assuming this powering mechanism. Here, we provide the setup of the explosions to incorporate the Ni--Co decay model. The setups for simulating the synthetic explosion using {\tt SNEC} and {\tt STELLA} closely follow \citet[][]{Ouchi2019} and \citet[][]{Aryan2021}, respectively. Here, we briefly summarise the important parameters and modifications made to \citet[][]{Ouchi2019} and \citet[][]{Aryan2021}.

We simulate the synthetic explosions of the 9\,M$_{\odot}$ model using {\tt SNEC}. First, the innermost 1.4\,M$_{\odot}$ is excised before the explosion, assuming that the model collapses to form neutron stars. The number of grid cells is set to be 1000 so that the light curves and photospheric velocities of the SNe from synthetic explosions are well converged in the time domain of interest.

For the M9.0\_Z0.0200\_Mni0.034\_E0.56 model, the synthetic explosion is carried out using {\tt SNEC}. The explosion is simulated as a {\tt Thermal\_Bomb} by adding $0.56 \times 10^{51}$\,erg of energy in the inner 0.1\,M$_{\odot}$ of the model for a duration of 0.1\,s. {\tt SNEC} does not include a nuclear-reaction network, so the amount of $^{56}$Ni is set by hand. A total of 0.034\,M$_{\odot}$ of $^{56}$Ni is distributed from the inner boundary up to the mass coordinate $m(r) = 2.0$\,M$_{\odot}$.

For the models having ZAMS masses of 12, 13, and 14\,M$_{\odot}$, we used {\tt STELLA} to simulate the synthetic explosions. The pre-SN model masses from 12\,M$_{\odot}$ models lie in the range of (3.45--3.96)\,M$_{\odot}$, while from 13\,M$_{\odot}$ models, the pre-SN model masses lie in the range (3.79--3.81)\,M$_{\odot}$. Furthermore, the 14\,M$_{\odot}$ model has a pre-SN mass of 4.23\,M$_{\odot}$ and is thus prone to produce a much higher $M_{\rm ej}$ than required for SN~2016iyc. For producing the synthetic explosions, the hydrodynamic simulations are performed using {\tt Thermal\_Bomb}-type explosion. Various explosion parameters including the ejecta masses, synthesised nickel masses, and explosion energies corresponding to different models are presented in Table~\ref{tab:MESA_MODELS}

The results of the hydrodynamic simulations are shown in Figure~\ref{fig:hydrodynamical_lc}. The left panel of Figure~\ref{fig:hydrodynamical_lc} shows the comparison of the {\tt SNEC}- and {\tt STELLA}-calculated bolometric light curves with the observed bolometric light curve (see Sec.~\ref{subsec3.3} for details on bolometric light curves) produced by fitting black bodies to the SEDs and integrating the fluxes over the wavelength range of 100--25,000\,\AA, while the right-hand panel shows the comparison of the corresponding photospheric velocities with the photospheric velocity obtained using the only available spectrum indicated by the Fe~II line velocity. The M9\,M\_Z0.0200\_Mni\_0.034\_E0.56 model could match the observed stretch factor and peak of the bolometric light curve, but it fails to reproduce the early extended SBO feature. The failure in producing the generic extended-SBO feature could be associated to the compactness of the pre-SN model having a radius of only 0.14\,R$_{\odot}$.

Moreover, all of the remaining models generate the generic extended-SBO feature, but only the M12.0\,M\_Z0.0200\_Mni0.025\_E0.35 could generate the extended-SBO and overall bolometric light curve that could match with actual bolometric light curve of SN~2016iyc. Another model that could nearly match the SN~2016iyc bolometric light curve is M13.0\,M\_Z0.0200\_Mni0.024\_E0.28. The remaining models deviate largely from the observed bolometric light curve of SN~2016iyc (left panel, Figure~\ref{fig:hydrodynamical_lc}). From the right-hand panel of Figure~\ref{fig:hydrodynamical_lc}, we find that the photospheric velocity evolution generated by the models, M13.0\,M\_Z0.0200\_Mni0.01\_E0.25 and M12.0\,M\_Z0.0200\_Mni0.025\_E0.35 pass closely to the observed line velocity from Fe~II which is a good indicator of observed photospheric velocity.

We also have an upper limit on the bolometric luminosity of SN~2016iyc at a phase nearly 220\,d since explosion. To produce the luminosity at that epoch, the model M12.0\,M\_Z0.0200\_Mni0.025\_E0.35 requires $M_{\rm Ni} = 0.09$\,M$_{\odot}$ (inset plot in the left panel of Figure~\ref{fig:hydrodynamical_lc}; M12.0\,M\_Z0.0200\_Mni0.09\_E0.35 is the corresponding model), serving as an upper limit on the synthesised nickel in SN~2016iyc.

\citet[][]{Anderson2019} has estimated a median value of 0.102\,M$_{\odot}$ for the nickel mass by considering 27 SNe~IIb. Moreover, \citet[][]{Afsariardchi2019} have also estimated the Ni mass for eight SNe~IIb and found that except for SN\,1996cb ($M_{\rm Ni} = 0.04\pm0.01$\,M$_{\odot}$) and SN\,2016gkg ($M_{\rm Ni} = 0.09\pm0.02$\,M$_{\odot}$), each SN~IIb has higher $M_{\rm Ni}$ than 0.09\,M$_{\odot}$. These comparisons show that SN~2016iyc definitely suffered low nickel production. Thus, the one-dimensional stellar evolution of various models along with the hydrodynamic simulations of their explosions suggest that a ZAMS progenitor having mass in the range 12--13\,M$_{\odot}$ with $M_{\rm ej}$ in the range (1.89--1.93)\,M$_{\odot}$, $M_{\rm Ni} < 0.09$\,M$_{\odot}$, and $E_{\rm exp}$ = (0.28--0.35) $\times 10^{51}$\,erg could be the possible progenitor of SN~2016iyc.

\begin{figure*}
\centering
    \includegraphics[height=8.0cm,width=8.5cm,angle=0]{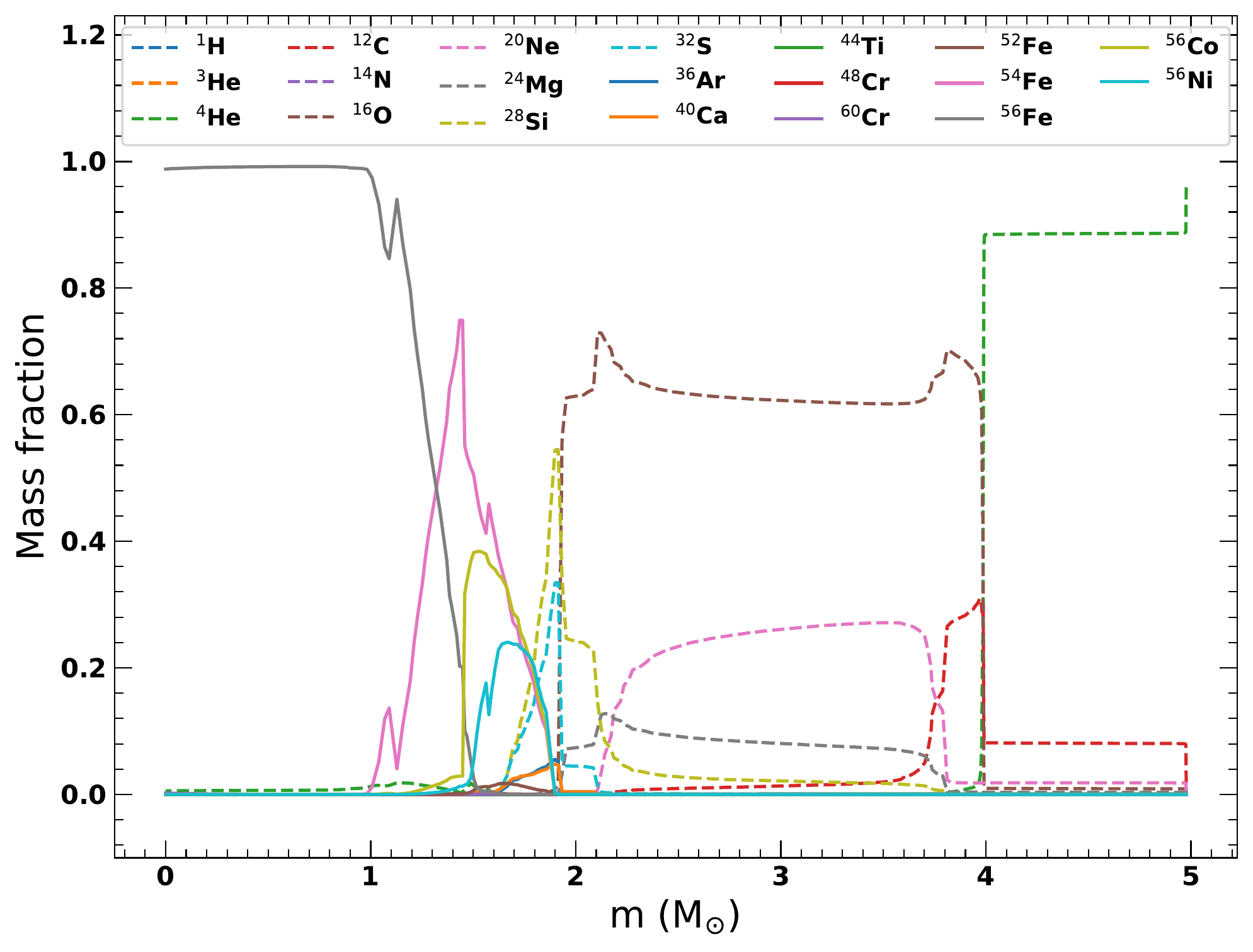}
    \includegraphics[height=8.0cm,width=8.5cm,angle=0]{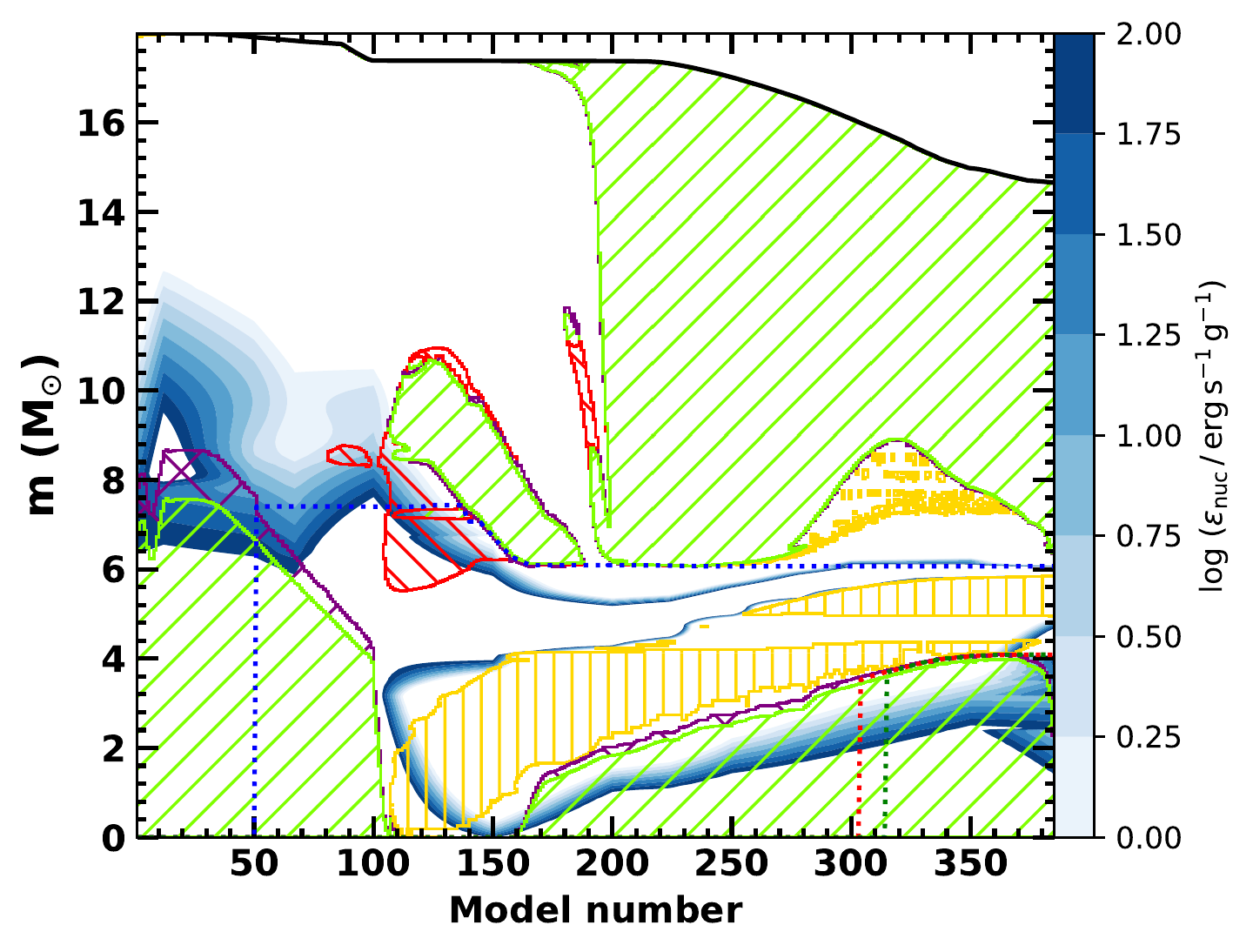}
   \caption{ {\em Left:} The mass fractions of a few key elements when the 18\,M$_{\odot}$ ZAMS progenitor model with $Z = 0.02$ has just reached the stage of Fe-core infall. The mass fraction of $^{56}$Fe in the centre is much higher compared to other species. {\em Right:} The Kippenhahn diagram of the model for a period from the beginning of main-sequence evolution to the stage when the model is ready to be stripped.}
    \label{fig:mass_Kipp_2}
\end{figure*}

\begin{figure*}
\centering
    \includegraphics[height=8.0cm,width=8.5cm,angle=0]{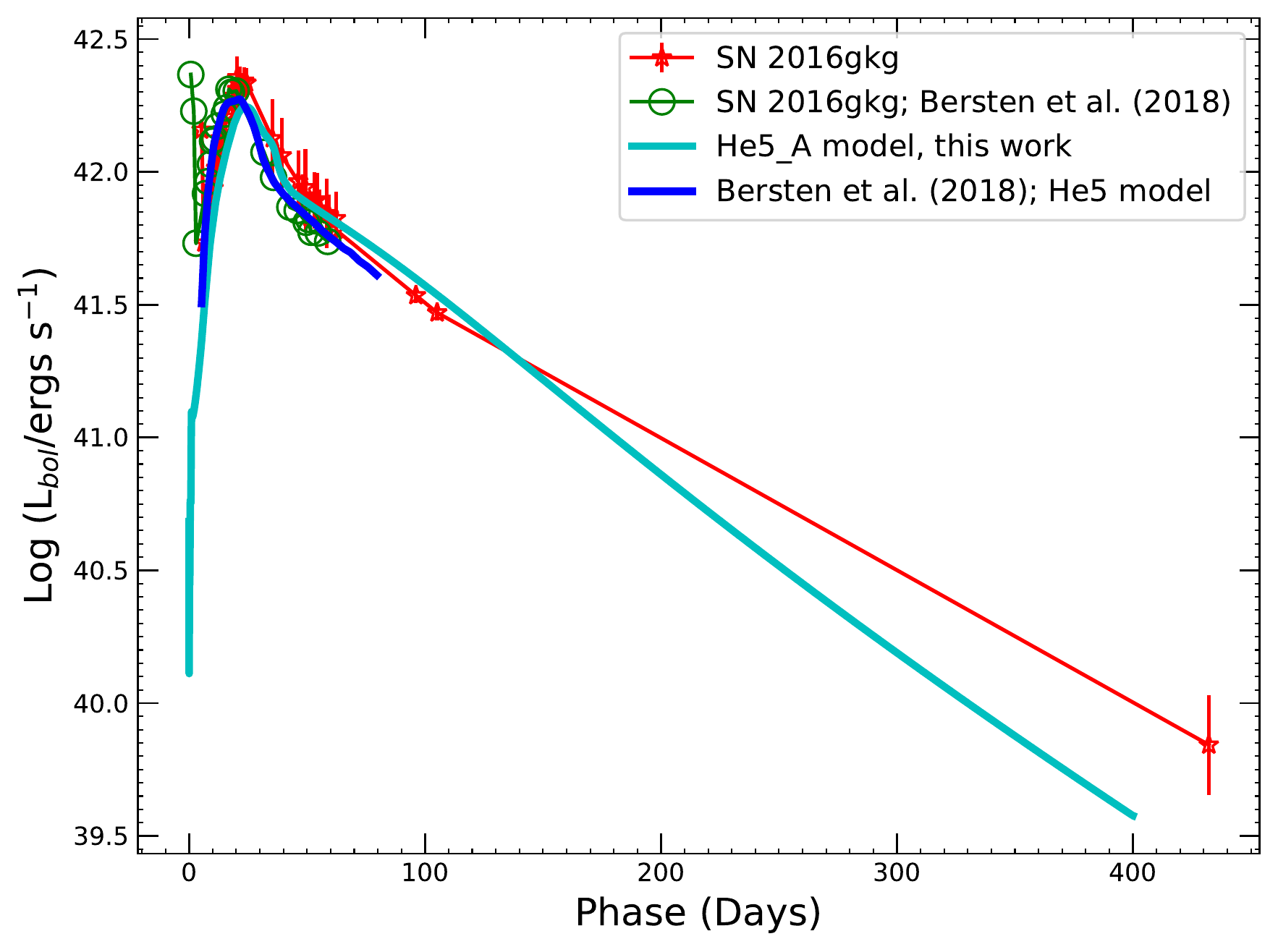}
    \includegraphics[height=8.0cm,width=8.5cm,angle=0]{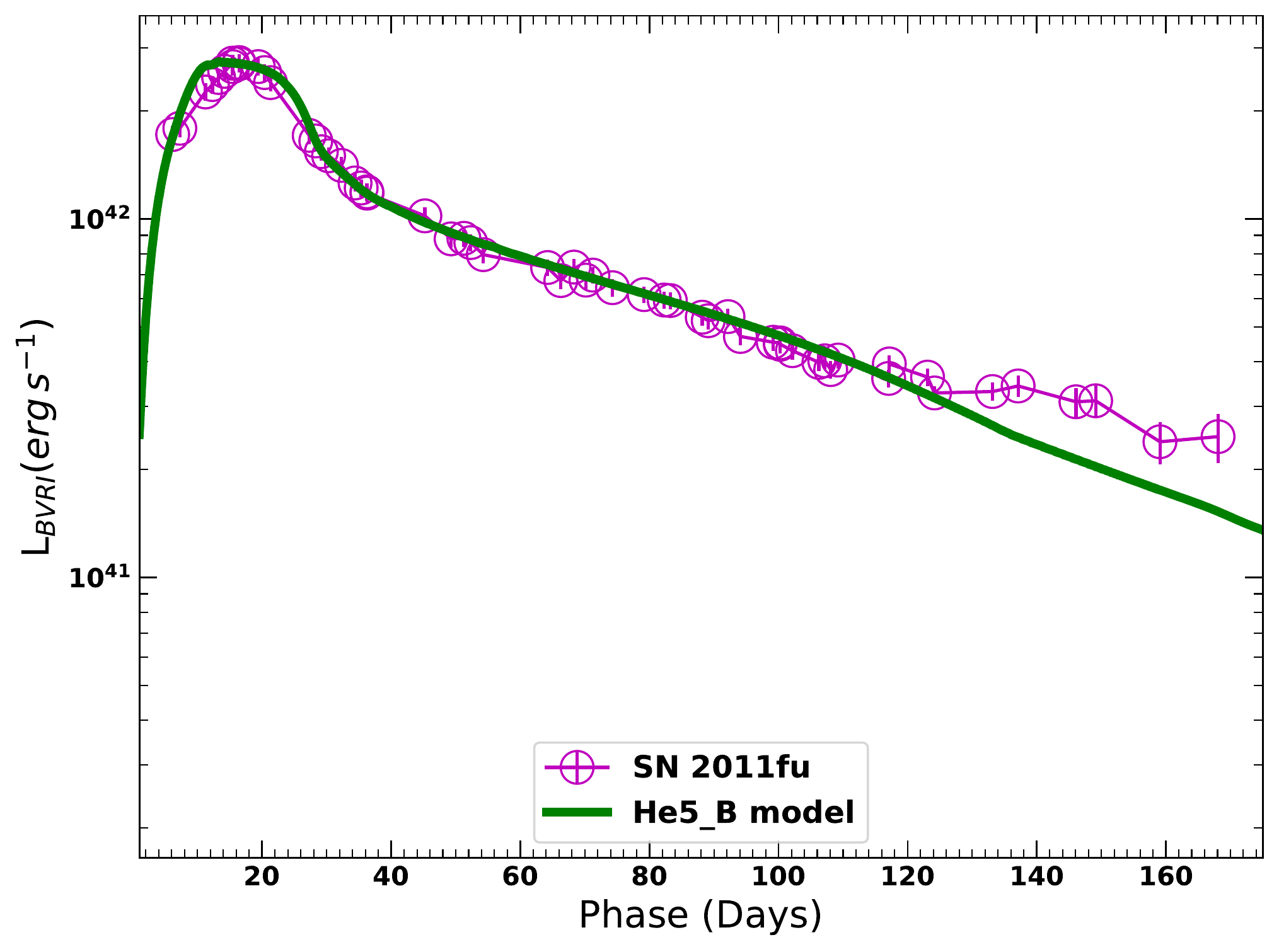}
   \caption{ The results of one-dimensional stellar evolution of models using {\tt MESA} and their synthetic explosion using {\tt SNEC} for SN~2016gkg and SN~2011fu. {\em Left:} Comparison of the quasibolometric light curve of SN~2016gkg with that obtained using {\tt SNEC} by taking into account the $^{56}$Ni and $^{56}$Co decay model and keeping the parameters close to those of \citet[][]{Bersten2018}. {\em Right:} Result of a similar analysis for SN~2011fu.}
    \label{fig:snec_2}
\end{figure*}

Recent studies suggest the masses of possible progenitors of Type IIb CCSNe to be usually higher than 9\,M$_{\odot}$, lying in the range 10--18\,M$_{\odot}$ \citep[][]{Van2013, Folatelli2014, Smartt2015}. However, there has been no direct observational evidence of an SN~IIb arising from a ZAMS progenitor of $\lesssim 12$\,M$_{\odot}$. The present analysis indicates that SN~2016iyc arises from the lower-mass end of the SN~IIb progenitor channel. As part of our study, we also performed the one-dimensional stellar evolutions of the possible progenitors of SN~2016gkg and SN~2011fu, and we simulated their hydrodynamic explosions in the next section to cover the range of faintest (SN~2016iyc), intermediate (SN~2016gkg), and highest (SN~2011fu) luminosity SNe in the comparison sample.

\section{Stellar modelling and synthetic explosions for SN 2016gkg and SN 2011fu}
\label{sec:SN2016gkg_model}

In this section, we perform hydrodynamic simulations of explosions from the possible progenitors of an intermediate-luminosity SN~2016gkg and the most-luminous SN~2011fu in the comparison sample to cover the higher end of the progenitor masses of SNe~IIb. After modelling their progenitors using {\tt MESA}, we simulate the synthetic explosions using {\tt SNEC} and match the {\tt SNEC} produced bolometric light curves with the observed ones. 

To construct the bolometric light curve of SN~2016gkg, we used the recalibrated $BVRI$ KAIT data along with the data from the 3.6\,m DOT at two epochs and incorporated {\tt SUPERBOL}. The photometric data of SN~2016gkg in this work are presented in Table~\ref{tab:optical_observations_2016gkg}. Previously, \citet[][]{Bersten2018} also used KAIT data calibrated from an older KAIT reduction pipeline. Figure~\ref{fig:SN2016gkg_comparison} shows the comparison between the KAIT data used by \citet[][]{Bersten2018} and the recalibrated KAIT data. To construct the bolometric light curve of SN~2011fu, we make use of {\tt SUPERBOL} as we did in earlier sections by incorporating the $BVRI$ data from \citet[][]{Kumar2013}. 

To model the possible progenitor of SN~2016gkg, we closely follow the HE5 model from \citet[][]{Bersten2018}. Also, \citet[][]{Morales2015} suggests a nearly similar model for the possible progenitor of SN~2011fu. An 18\,M$_{\odot}$ ZAMS progenitor mass is employed for both SNe. The modelling and explosion parameters are listed in Table~\ref{tab:MESA_MODELS_11fu_n_16gkg}. Starting from the ZAMS, the model is evolved up to the stage where the core starts to collapse. The evolution of the model on the HR diagram is shown in the left panel of Figure~\ref{fig:HR_Rho_T}. Various physical processes during the evolution on the HR diagram have been indicated. Also, the right-hand panel displays the variation of $T_{\rm core}$ with $\rho_{\rm core}$. It is indicated that during the last evolutionary phases, the core density and temperatures have reached over $10^{10}$\,g\,cm$^{-3}$ and $10^{10}$\,K, respectively. Such high core density and temperatures mark the onset of core collapse. The left panel of Figure~\ref{fig:mass_Kipp_2} shows the mass fractions of various elements at the stage when the model has just reached the stage of Fe-core infall. As another piece of evidence for the onset of core collapse, we can see that the core is mainly composed of inert $^{56}$Fe. The right-hand panel of Figure~\ref{fig:mass_Kipp_2} shows the Kippenhahn diagram of the model for a period from the beginning of main-sequence evolution to the stage when the model is ready to be stripped.

Models He5\_A and He5\_B are used for SN~2016gkg and SN~2011fu, respectively. Although the parameters including the ZAMS mass, metallicity, rotation, and overshoot parameter are same for these two models, different explosion parameters are employed using {\tt SNEC} to simulate the synthetic explosions.

The left panel of Figure~\ref{fig:snec_2} illustrates the comparison of our hydrodynamic simulation of synthetic explosions for SN~2016gkg with the results of \citet[][]{Bersten2018}. The difference between the bolometric light curve from \citet[][]{Bersten2018} and calculated using KAIT revised photometry are within the error bars. Our model could explain the bolometric light curve of SN~2016iyc very well. Furthermore, the right-hand panel of Figure~\ref{fig:snec_2} shows the comparison of the {\tt SNEC}-calculated bolometric light curve with the observed quasibolometric light curve of SN~2011fu. The one-dimensional stellar modelling of possible progenitors using {\tt MESA} along with their hydrodynamic simulation of explosions using {\tt SNEC} explain the observed light curves of SN~2016gkg and SN~2011fu very well. Now, we have performed the stellar modelling of the possible progenitors and the hydrodynamic explosions of SN~2016iyc, SN~2016gkg, and SN~2011fu to cover a range of faintest (SN~2016iyc), intermediate (SN~2016gkg), and highest (SN~2011fu) luminosity SNe in the comparison sample.

\section{Discussion}
\label{sec:Discussions}

Detailed photometric and spectroscopic analyses of the low-luminosity Type IIb SN~2016iyc are performed in this work. The extinction-corrected data of SN~2016iyc are used to construct the quasibolometric and bolometric light curves using {\tt SUPERBOL}. Comparisons of the absolute $V$-band and quasibolometric light curves of SN~2016iyc with other well-studied SNe~IIb indicates that SN~2016iyc lies toward the faint limit of this subclass. Low-luminosity SNe~IIb with low $^{56}$Ni production are thought to arise from progenitors having masses near the threshold mass for producing a CCSN.

Our study indicates that among the comparison sample in this work, SN~2016iyc has the smallest black-body radius at any given epoch. This anomalous behaviour could be attributed to its low ejecta velocity.

Based on the low $M_{\rm ej}$ and the lowest intrinsic brightness among SNe in the comparison sample, 9--14\,M$_{\odot}$ ZAMS progenitors are modelled as the possible progenitor of SN~2016iyc using {\tt MESA}. The results of synthetic explosions simulated using {\tt STELLA} and {\tt SNEC} are in good agreement with the observed ones.  

The one-dimensional stellar modelling of the possible progenitor using {\tt MESA} and simulations of hydrodynamic explosions using {\tt SNEC}/{\tt STELLA} indicate that SN~2016iyc originated from a (12--13)\,M$_{\odot}$ ZAMS progenitor, near the lower end of progenitor masses for SNe~IIb. The models show a range of parameters for SN~2016iyc, including $M_{\rm ej} =$ (1.89--1.93)\,M$_{\odot}$ and $E_{\rm exp} =$ (0.28--0.35) $\times 10^{51}$\,erg. We also put an upper limit of 0.09\,M$_{\odot}$ on the amount of nickel synthesised by the SN. The pre-SN radius of the progenitor of SN~2016iyc is (240--300)\,R$_{\odot}$.

Stellar evolution of the possible progenitors and hydrodynamic simulations of synthetic explosions of SN~2016gkg and SN~2011fu have also been performed to compare the intermediate- and high-luminosity ends among well-studied SNe~IIb using {\tt MESA} and {\tt SNEC}. The results of stellar modelling and synthetic explosions for SN~2016iyc, SN~2016gkg, and SN~2011fu exhibit a diverse range of mass of the possible progenitors for SNe~IIb.

\section{Conclusions}
\label{sec:Conclusions}

We performed detailed photometric and spectroscopic analyses of SN~2016iyc, a Type IIb SN discovered by LOSS/KAIT. The observed photometric properties of SN~2016iyc were unique in many ways: low luminosity, low ejecta mass, and small black-body radius.  Attempts to model the possible progenitor were made using the one-dimensional hydrodynamic code {\tt MESA}. As a part of the present work,  hydrodynamic modelling of the synthetic explosion of an intermediate-luminosity Type IIb SN~2016gkg using recalibrated KAIT data and late-time data from the 3.6\,m DOT, along with an optically very luminous Type IIb SN~2011fu, were also performed. The main results based on the present analysis are as follows.

\begin{enumerate}

\item{Based on the low value of $M_{\rm ej}$, ZAMS stars having masses of 9--14\,M$_{\odot}$ were adopted to model the possible progenitor of SN~2016iyc using {\tt MESA}. The results of synthetic explosions simulated using {\tt SNEC} and {\tt STELLA} were in good agreement with observed properties for ZAMS progenitor masses of 12--13\,M$_{\odot}$ having a pre-SN radius of (240--300)\,R$_{\odot}$. Thus, SN~2016iyc likely had a progenitor arising from the lower end of the progenitor mass channel of an SN~IIb.}\\

\item{We concluded that the overall detailed hydrodynamic simulations of the explosions from various models showed a range of parameters for SN~2016iyc, including an $M_{\rm ej}$ of (1.89--1.93)\,M$_{\odot}$, an $E_{\rm exp}$ of (0.28--0.35) $\times 10^{51}$\,erg, and an upper limit of  $< 0.09$\,M$_{\odot}$ on the amount of nickel synthesised by SN~2016iyc.}\\   
 
\item{Finally, one-dimensional stellar evolution models of possible progenitors and the hydrodynamic explosions of SN~2016gkg and SN~2011fu were also performed to compare intermediate- and high-luminosity examples among well-studied SNe~IIb. The results for SN~2016iyc, SN~2016gkg, and SN~2011fu showed a diverse range of mass [(12.0--18.0)\,M$_{\odot}$] for the possible progenitors of SNe~IIb considered in this work. Discovery of more such events through survey projects in the near future should provide additional data with which to establish the lower mass limits of such explosions.}
\end{enumerate}

\section*{Acknowledgements}
We thank the anonymous referee for providing very useful and constructive comments that helped to improve the manuscript significantly. A.A. acknowledges funds and assistance provided by the Council of Scientific \& Industrial Research (CSIR), India with the file no. 09/948(0003)/2020-EMR-I. A.A., S.B.P., and R.G. also acknowledge BRICS grant DST/IMRCD/BRICS/Pilotcall/ProFCheap/2017(G). RG and SBP acknowledge the financial support of ISRO under AstroSat archival Data utilization program (DS$\_$2B-13013(2)/1/2021-Sec.2). We sincerely acknowledge the extensive use of the High Performance Computing (HPC) facility at ARIES.

Support for A.V.F.'s supernova research group has been provided by the TABASGO Foundation, the Christopher R. Redlich Fund, the U.C. Berkeley Miller Institute for Basic Research in Science (where A.V.F. was a Senior Miller Fellow), and numerous individual donors. Additional support was provided by NASA/{\it HST} grant GO-15166 from the Space Telescope Science Institute (STScI), which is operated by the Associated Universities for Research in Astronomy, Inc. (AURA), under NASA contract NAS 5-26555. J.V. is supported by the project ``Transient Astrophysical Objects'' (GINOP 2.3.2-15-2016-00033) of the National Research, Development, and Innovation Office (NKFIH), Hungary, funded by the European Union. We acknowledge Prof. Keiichi Maeda for useful scientific discussions. The "Open Supernova Catalog" is duly acknowledged here for spectroscopic data.

Lick/KAIT and its ongoing operation were made possible by donations from Sun Microsystems, Inc., the Hewlett-Packard Company, AutoScope Corporation, Lick Observatory, the U.S. National Science Foundation, the University of California, the Sylvia \& Jim Katzman Foundation, and the TABASGO Foundation. Research at Lick Observatory is partially supported by a generous gift from Google. Some of the data presented herein were obtained at the W. M. Keck Observatory, which is operated as a scientific partnership among the California Institute of Technology, the University of California, and NASA; the observatory was made possible by the generous financial support of the W. M. Keck Foundation. The Lick and Keck Observatory staff provided excellent assistance with the observations. 

This work makes use of observations from the Las Cumbres Observatory Global Telescope Network. The authors sincerely acknowledge M. Garcia and R. Case for performing the Nickel observations.

\section*{Data availability}
The photometric and spectroscopic data used in this work, as well as the {\tt inlist} files to create the {\tt MESA} models, can be made available on reasonable request to the corresponding author.








\appendix
\section{Tables and figures}

\begin{table*}
\caption{Photometry of SN~2016iyc}
\centering
\smallskip
\begin{tabular}{c c c c c c c c}
\hline \hline
MJD  	    &  $B$              &  $V$          &  $R$    & $C$      &  $I$        & Telescope               \\
         &(mag)		    & (mag)             & (mag)         & (mag)         & (mag)                     \\
\hline                           

57740.144	& ...	& ...  &  ...	& 17.807 $\pm$	 0.111   	& ...  & KAIT \\
57741.116	& 18.473 $\pm$	 0.075	& 18.262 $\pm$	 0.074	& 18.081 $\pm$	 0.091	& 17.985 $\pm$	 0.048   	& 17.875 $\pm$	 0.103  & KAIT \\
57742.089	& 18.774 $\pm$	 0.149	& 18.484 $\pm$	 0.118	& 18.102 $\pm$	 0.147	& 18.057 $\pm$	 0.149   	& 17.832 $\pm$	 0.148  & KAIT \\
57743.109	& 18.875 $\pm$	 0.114	& 18.484 $\pm$	 0.111	& 18.105 $\pm$	 0.117	& 17.987 $\pm$	 0.135	    & 17.949 $\pm$	 0.142  & KAIT \\
57744.087	& 18.796 $\pm$	 0.139	& 18.451 $\pm$	 0.113	& 17.957 $\pm$	 0.113	& 17.957 $\pm$	 0.143	    & 17.731 $\pm$	 0.126  & KAIT \\
57744.091	& 18.787 $\pm$	 0.054	& 18.299 $\pm$	 0.050	& 17.933 $\pm$	 0.067	& -- $\pm$	 --	    & 17.656 $\pm$	 0.119  & Nickel \\
57745.092	& 18.865 $\pm$	 0.129	& 18.237 $\pm$	 0.084	& 18.038 $\pm$	 0.106	& 17.855 $\pm$	 0.115	    & 17.658 $\pm$	 0.114  & KAIT \\
57749.111	& 18.568 $\pm$	 0.105	& 18.090 $\pm$	 0.081	& 17.728 $\pm$	 0.098	& 17.742 $\pm$	 0.131	    & 17.524 $\pm$	 0.130  & KAIT \\
57750.090	& 18.582 $\pm$	 0.100	& 18.050 $\pm$	 0.075	& 17.735 $\pm$	 0.080	& 17.693 $\pm$	 0.092	    & 17.433 $\pm$	 0.084  & KAIT \\
57751.095	& 18.567 $\pm$	 0.085	& 18.037 $\pm$	 0.063	& 17.693 $\pm$	 0.070	& 17.665 $\pm$	 0.055      & 17.439 $\pm$	 0.073  & KAIT \\ 
57751.095	& 18.567 $\pm$	 0.085	& 18.037 $\pm$	 0.063	& 17.693 $\pm$	 0.070	& 17.665 $\pm$	 0.055      & 17.439 $\pm$	 0.073  & KAIT \\
57753.099	& 18.578 $\pm$	 0.083	& 18.009 $\pm$	 0.061	& 17.679 $\pm$	 0.063	& 17.635 $\pm$	 0.091      & 17.408 $\pm$	 0.082  & KAIT \\
57754.091	& 18.724 $\pm$	 0.137	& 17.947 $\pm$	 0.066	& 17.678 $\pm$	 0.067	& 17.653 $\pm$	 0.090      & 17.409 $\pm$	 0.093  & KAIT \\
57768.107	& ...	& 18.638 $\pm$	 0.059	& 18.582 $\pm$	 0.395	& 18.570 $\pm$	 0.242      & 18.069 $\pm$	 0.321  & KAIT \\
57769.105	& 19.295 $\pm$	 0.464	& 19.111 $\pm$	 0.214	& 18.666 $\pm$	 0.194	& 18.784 $\pm$	 0.282      & 18.466 $\pm$	 0.232  & KAIT \\
57956.471	& $>$21.490 	& $>$21.336	& $>$21.542 	& ...       & $>$20.631   & Nickel\\
\hline                                   
\end{tabular}
\label{tab:optical_observations_2016iyc}      
\end{table*}

\begin{table*}
\caption{Revised KAIT photometry of SN~2016gkg along with 3.6\,m DOT data}
\centering
\smallskip
\begin{tabular}{c c c c c c c c}
\hline \hline
MJD  	    &  $B$              &  $V$          &  $R$    & $C$      &  $I$        & Telescope               \\
         &(mag)		    & (mag)             & (mag)         & (mag)         & (mag)                     \\
\hline                           

57653.315	& 15.949 $\pm$	 0.357	& 15.799 $\pm$	 0.234	& 15.642 $\pm$	 0.214	& 15.661 $\pm$	 0.069   	& 15.518 $\pm$	 0.286  & KAIT \\
57654.322	& 17.039 $\pm$	 0.079	& 16.569 $\pm$	 0.113	& 16.257 $\pm$	 0.053	& 16.272 $\pm$	 0.030   	& 16.056 $\pm$	 0.068  & KAIT \\
57658.373	& 16.626 $\pm$	 0.054	& 15.990 $\pm$	 0.038	& 15.696 $\pm$	 0.045	& 15.745 $\pm$	 0.045   	& 15.652 $\pm$	 0.050  & KAIT \\
57659.444	& 16.383 $\pm$	 0.049	& 15.854 $\pm$	 0.041	& 15.557 $\pm$	 0.053	& 15.630 $\pm$	 0.042	    & 15.539 $\pm$	 0.063  & KAIT \\
57660.444	& 16.282 $\pm$	 0.047	& 15.690 $\pm$	 0.018	& 15.450 $\pm$	 0.022	& 15.429 $\pm$	 0.083	    & 15.401 $\pm$	 0.045  & KAIT \\
57661.316	& 16.106 $\pm$	 0.066	& 15.566 $\pm$	 0.052	& 15.249 $\pm$	 0.061	& 15.482 $\pm$	 0.071	    & 15.183 $\pm$	 0.079  & KAIT \\
57662.408	& 16.024 $\pm$	 0.060	& 15.466 $\pm$	 0.053	& 15.213 $\pm$	 0.059	& 15.238 $\pm$	 0.062	    & 15.182 $\pm$	 0.085  & KAIT \\
57663.334	& 15.861 $\pm$	 0.074	& 15.378 $\pm$	 0.068	& 15.125 $\pm$	 0.162	& 15.145 $\pm$	 0.068	    & 15.049 $\pm$	 0.109  & KAIT \\
57666.368	& 15.748 $\pm$	 0.039	& 15.217 $\pm$	 0.034	& 15.013 $\pm$	 0.045	& 14.992 $\pm$	 0.033	    & 14.887 $\pm$	 0.049  & KAIT \\
57667.373	& 15.676 $\pm$	 0.037	& 15.162 $\pm$	 0.038	& 14.969 $\pm$	 0.051	& 14.931 $\pm$	 0.025      & 14.868 $\pm$	 0.050  & KAIT \\ 
57668.362	& 15.504 $\pm$	 0.035	& 15.009 $\pm$	 0.012	& 14.838 $\pm$	 0.011	&   ...       & 14.763 $\pm$	 0.013  & KAIT \\
57668.375	& 15.605 $\pm$	 0.117	& 15.093 $\pm$	 0.060	& 14.916 $\pm$	 0.132	& 14.830 $\pm$	 0.084      & 14.805 $\pm$	 0.146  & KAIT \\
57669.369	& 15.565 $\pm$	 0.117	& 15.011 $\pm$	 0.139	& 14.845 $\pm$	 0.128	& 14.758 $\pm$	 0.180      & 14.671 $\pm$	 0.024  & KAIT \\
57671.420	& 15.536 $\pm$	 0.058	& 15.024 $\pm$	 0.043	& 14.760 $\pm$	 0.051	& 14.759 $\pm$	 0.063      & 14.621 $\pm$	 0.069  & KAIT \\
57672.330	& 15.576 $\pm$	 0.048	& 15.003 $\pm$	 0.074	& 14.760 $\pm$	 0.072	& 14.723 $\pm$	 0.054      & 14.590 $\pm$	 0.090  & KAIT \\
57683.307	& 16.856 $\pm$	 0.023	& 15.699 $\pm$	 0.016	& 15.187 $\pm$	 0.014	&    ...   	& 14.951 $\pm$	 0.014  & KAIT \\
57687.298	& 17.165 $\pm$	 0.040	& 15.995 $\pm$	 0.017	& 15.383 $\pm$	 0.016	&     ...   	& 15.076 $\pm$	 0.016  & KAIT \\
57694.279	& 17.545 $\pm$	 0.096	& 16.266 $\pm$	 0.049	& 15.645 $\pm$	 0.054	& 15.692 $\pm$	 0.053	    & 15.226 $\pm$	 0.069  & KAIT \\
57696.255	& 17.544 $\pm$	 0.021  & 16.302 $\pm$	 0.016	& 15.716 $\pm$	 0.023	&   ...	    &  ...  & 3.6\,m DOT \\
57697.300	& 17.560 $\pm$	 0.863	& 16.379 $\pm$	 0.076	& 15.703 $\pm$	 0.020	&     ...	    & 15.363 $\pm$	 0.019  & KAIT \\
57697.350	& 17.562 $\pm$	 0.106	& 16.286 $\pm$	 0.106	& 15.697 $\pm$	 0.118	& 15.815 $\pm$	 0.016	    & 15.372 $\pm$	 0.120  & KAIT \\
57701.256	& 17.669 $\pm$	 0.081	& 16.440 $\pm$	 0.052	& 15.809 $\pm$	 0.059	& 15.891 $\pm$	 0.040	    & 15.387 $\pm$	 0.059  & KAIT \\
57702.253	& 17.617 $\pm$	 0.038	& 16.487 $\pm$	 0.018	& 15.833 $\pm$	 0.018	&     ...	    &     ...  & KAIT \\
57703.289	& 17.532 $\pm$	 0.108	& 16.463 $\pm$	 0.095	& 15.847 $\pm$	 0.071	& 15.884 $\pm$	 0.055	    & 15.426 $\pm$	 0.053  & KAIT \\
57706.262	& 17.805 $\pm$	 0.156	& 16.562 $\pm$	 0.071	& 15.894 $\pm$	 0.059	& 15.965 $\pm$	 0.086      & 15.543 $\pm$	 0.071  & KAIT \\ 
57707.237	& 17.592 $\pm$	 0.329	& 16.486 $\pm$	 0.232	& 15.829 $\pm$	 0.186	& 16.023 $\pm$	 0.037      & 15.469 $\pm$	 0.178  & KAIT \\
57710.259	& 17.893 $\pm$	 0.085	& 16.612 $\pm$	 0.055	& 16.011 $\pm$	 0.063	& 16.050 $\pm$	 0.046      & 15.535 $\pm$	 0.070  & KAIT \\
57710.312	& 16.692 $\pm$	 0.084	& 16.689 $\pm$	 0.038	& 16.077 $\pm$	 0.035	&     ...      & 15.617 $\pm$	 0.021  & KAIT \\ 
57744.149	& 18.032 $\pm$	 0.109	& 17.277 $\pm$	 0.061	& 16.678 $\pm$	 0.035	&     ...      & 16.143 $\pm$	 0.027  & KAIT \\
57753.135	& 18.208 $\pm$	 0.039	& 17.371 $\pm$	 0.089	& 16.854 $\pm$	 0.029	&     ...      & 16.282 $\pm$	 0.026  & KAIT \\
58080.168	& 22.541 $\pm$	 0.258    &  21.751 $\pm$	 0.225	& 20.821 $\pm$	 0.03	&   ...	    &  20.137 $\pm$	 0.055  & 3.6\,m DOT \\

\hline                                   
\end{tabular}
\label{tab:optical_observations_2016gkg}      
\end{table*}

\begin{table*}
\caption{{\tt MESA} model and {\tt SNEC} explosion parameters of SN~2011fu and SN~2016gkg.}
\label{tab:MESA_MODELS_11fu_n_16gkg}
\begin{center}
{\scriptsize
\begin{tabular}{ccccccccccccc} 
\hline\hline
SN name &Model Name	& $M_{\rm ZAMS}$	& $Z$	 & $f_{ov}^{a}$ &	$M_{\mathrm{f}}^{b}$	 & $M_\mathrm{c}^{c}$	 & $M_{\mathrm{ej}}^{d}$	&	$M_{\mathrm{Ni}}^{e}$ &	$E_{\mathrm{exp}}^{f}$ 	\\
&	&	(M$_{\odot}$)	&	 &  & (M$_{\odot}$)	&	(M$_{\odot}$)  & (M$_{\odot}$) & (M$_{\odot}$) & 	($10^{51}$\,erg) 	\\
\hline
\hline
SN~2016gkg & He5\_A  &	18.0  	&	0.0200      & 0.01   &    5.00  & 1.6 &  3.40  & 0.087  &  1.30		\\
SN~2011fu & He5\_B  &	18.0  	&	0.0200       & 0.01   &    5.00  & 1.5 &  3.50 & 0.140  &  1.25		\\
\hline\hline
\end{tabular}}
\end{center}
{$^a$Overshoot parameter.
$^b$Final mass of the pre-SN model.
$^c$Final mass of the central remnant
$^d$Ejecta mass.
$^e$Nickel mass
$^f$Explosion energy.}\\

\end{table*}

\renewcommand{\thefigure}{A\arabic{figure}}
\setcounter{figure}{0}

\begin{figure*}
\includegraphics[width=\columnwidth]{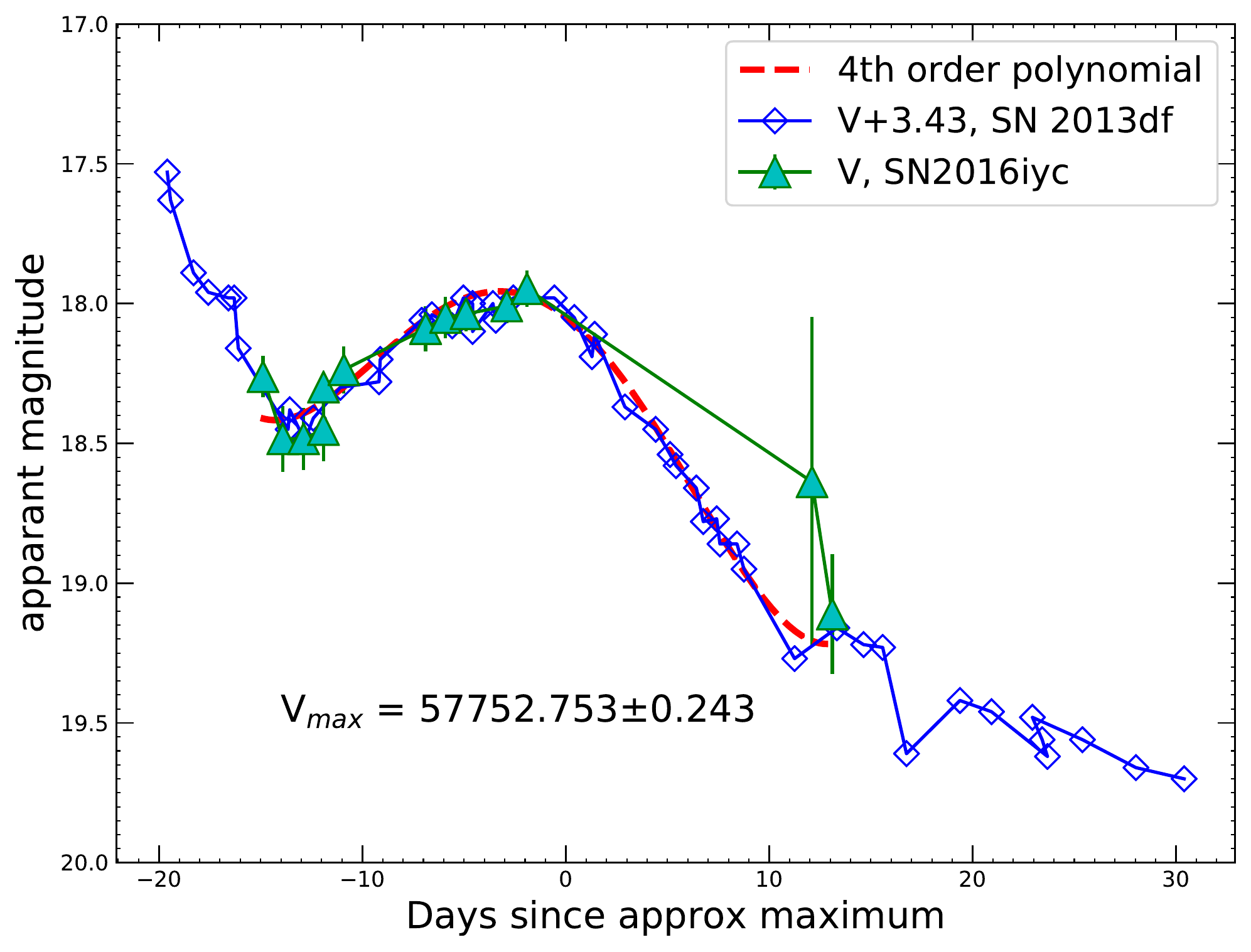}
\caption{Estimation of $V$-band maximum.}
\label{fig:V_max}
\end{figure*}

\begin{figure*}
\centering
    \includegraphics[height=7.0cm,width=8.8cm,angle=0]{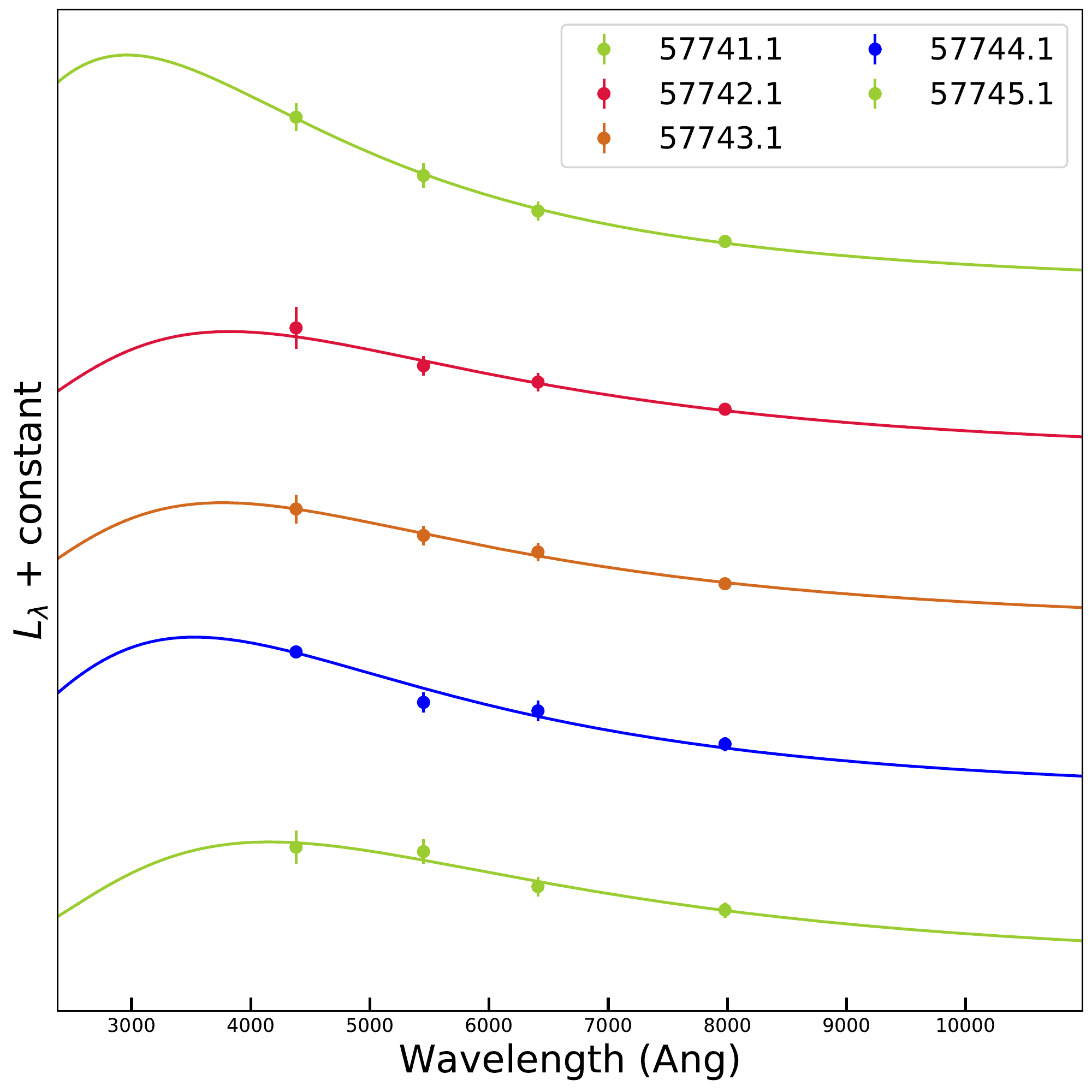}
    \includegraphics[height=7.0cm,width=8.5cm,angle=0]{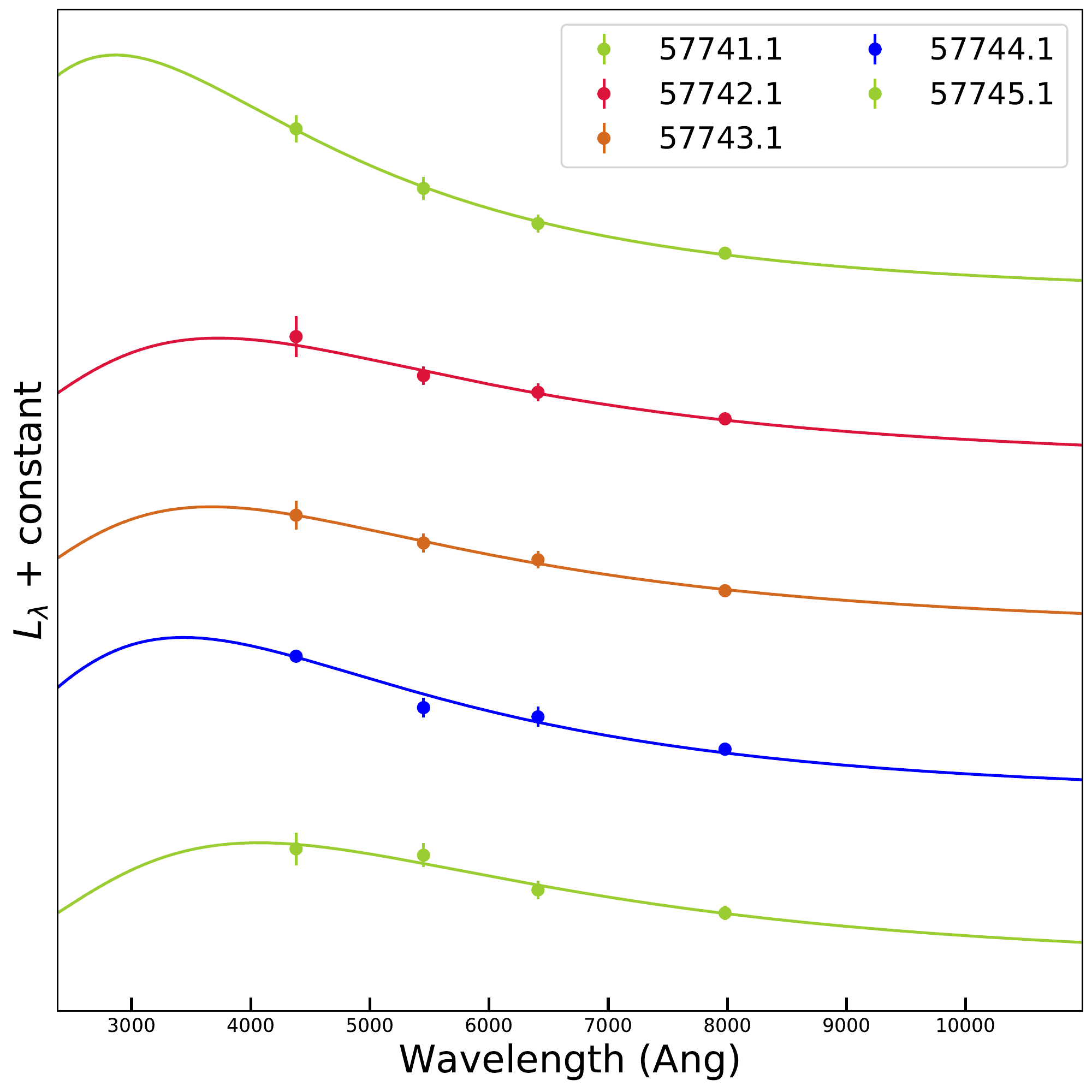}
    \includegraphics[height=7.0cm,width=8.5cm,angle=0]{bb_fits_iyc_0.00_BVRI.pdf}
    \includegraphics[height=7.0cm,width=8.5cm,angle=0]{bb_fits_iyc_0.02_BVRI.pdf}
    \includegraphics[height=7.0cm,width=8.5cm,angle=0]{bb_fits_iyc_0.00_BVRI.pdf}
    \includegraphics[height=7.0cm,width=8.5cm,angle=0]{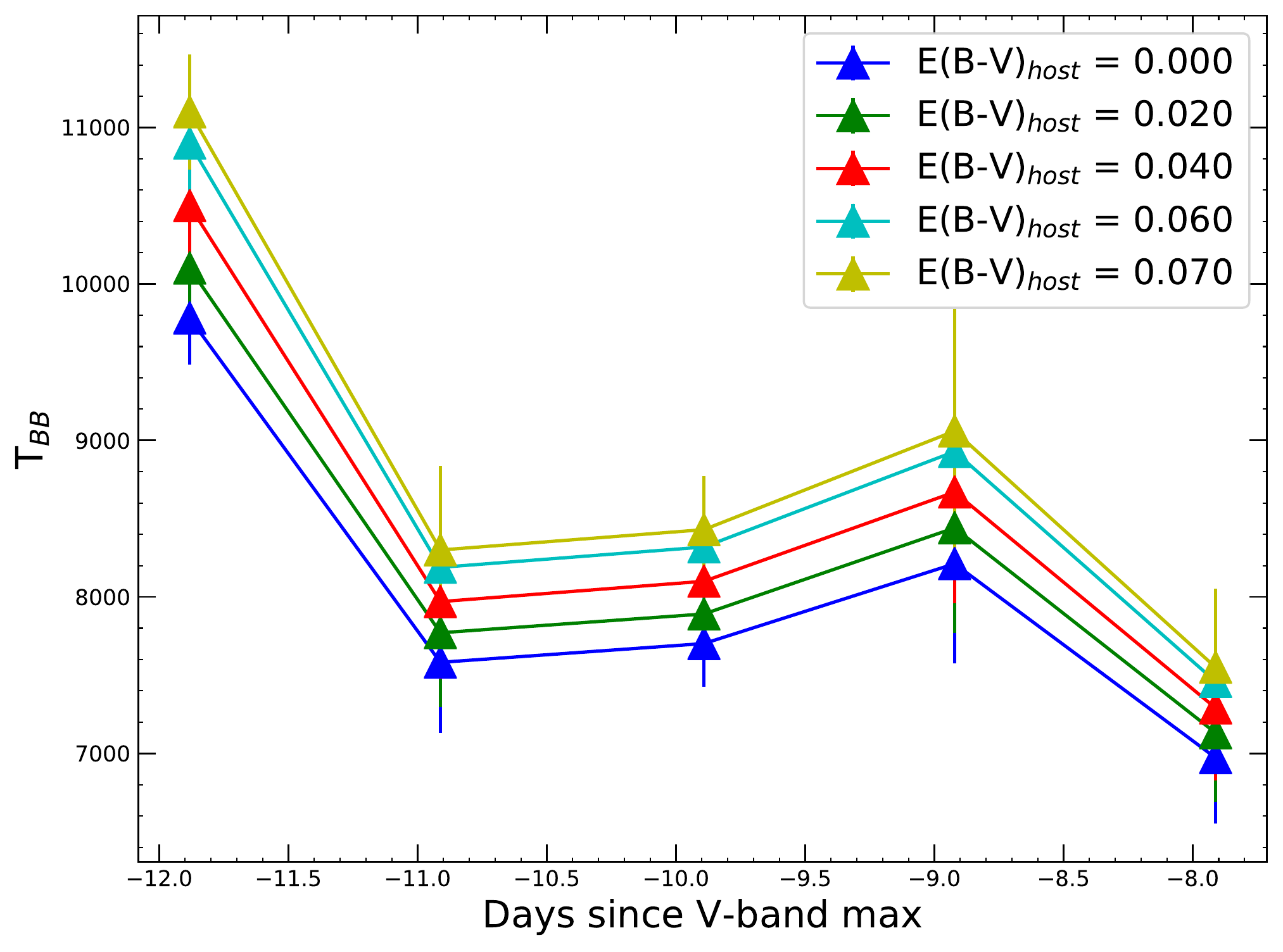}
   \caption{Fitting black-body curves to a few early epochs of SN~2016iyc by assuming different host-galaxy extinctions. The top-left and right panels show black-body fits to a few early-epoch SEDs of SN~2016iyc corresponding to host-galaxy extinctions of 0.00\,mag and 0.02\,mag, respectively. The middle-left and right panels show black-body fits to a few early-epoch SEDs of SN~2016iyc corresponding to host-galaxy extinctions of 0.04\,mag and 0.06\,mag, respectively. The bottom-left panel shows black-body fits to a few early-epoch SEDs of SN~2016iyc corresponding to a host-galaxy extinction of 0.07\,mag, while the bottom-right panel shows the variation of the black-body temperature obtained using black-body fits to the SEDs corresponding to different host-galaxy extinctions.}
    \label{fig:extinction_trial}
\end{figure*}

\begin{figure*}
\includegraphics[width=\columnwidth]{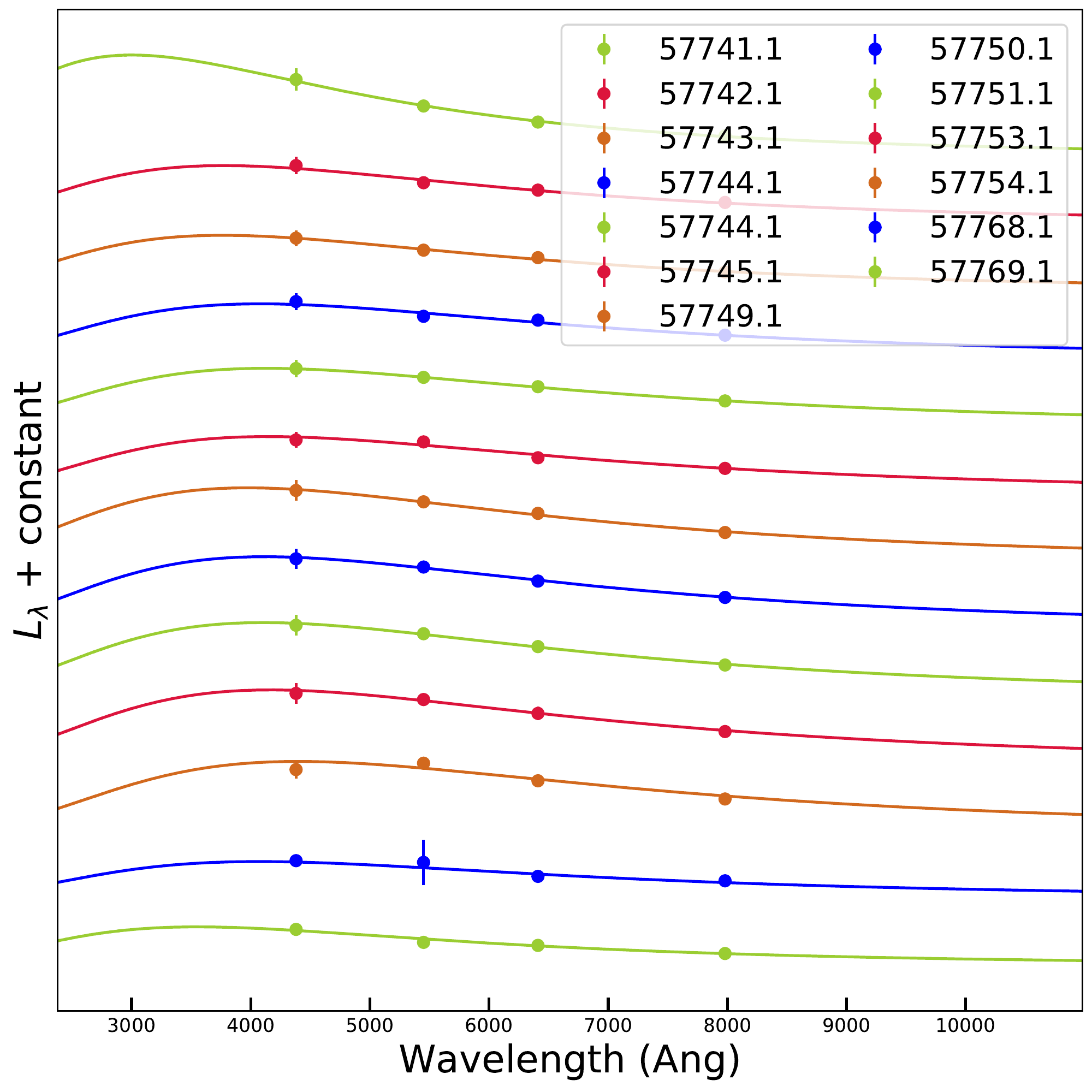}
\caption{The black-body fits to the SED of SN~2016iyc to estimate the bolometric light curve generated from {\tt SUPERBOL}.}
\label{fig:bb_fits}
\end{figure*}

\begin{figure*}
\includegraphics[width=\columnwidth]{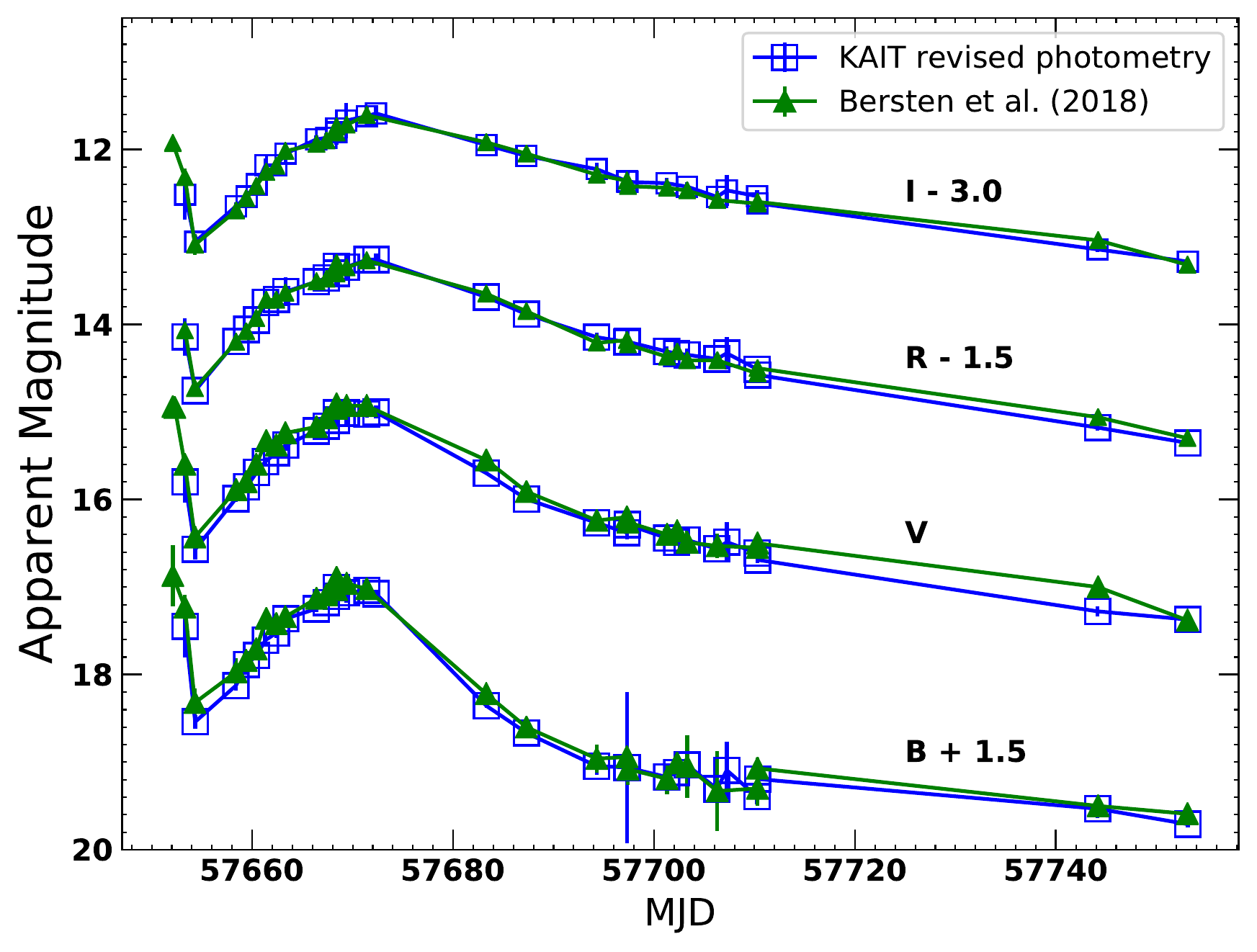}
\caption{Comparison between the KAIT revised photometry and the KAIT data used by \citet[][]{Bersten2018} for SN~2016gkg.}
\label{fig:SN2016gkg_comparison}
\end{figure*}



\bsp	
\label{lastpage}
\end{document}